\documentclass[10pt,journal,compsoc]{IEEEtran}
\ifCLASSOPTIONcompsoc
  \usepackage[nocompress]{cite}
\else
  \usepackage{cite}
\fi

\ifCLASSINFOpdf
\else
\fi

\usepackage{booktabs} 

\newcommand{\hide}[1]{}
\newcommand{\raf}[1]{(\ref{#1})}
\newcommand{\abs}[1]{\ensuremath{\left|#1\right|}}

\usepackage{graphicx}
\usepackage{algorithm,algpseudocode}
\usepackage{tabularx,adjustbox}
\usepackage{amsmath}

\usepackage{amsmath,amsfonts,amssymb,mathtools}

\usepackage{epsfig,amstext,xspace}

\usepackage{amsthm}

\usepackage{stfloats}
\usepackage[sans]{dsfont}
\usepackage{stmaryrd}
\usepackage{pifont}
\usepackage{wasysym}

\usepackage{array}
\usepackage{url}
\usepackage{color}
\usepackage{multirow}
\usepackage[skip=1pt]{subcaption}
\newcolumntype{x}[1]{>{\centering\arraybackslash\hspace{0pt}}m{#1}}

\newtheorem{theorem}{Theorem}

\begin{document}
%

\title{Cloud-based Privacy-Preserving Collaborative Consumption for Sharing Economy}

\author{Lingjuan Lyu, Sid Chi-Kin Chau, Nan Wang and  Yifeng Zheng  
\IEEEcompsocitemizethanks{
\IEEEcompsocthanksitem L. Lyu is with the Department of Computer Science, National University of Singapore. E-mail: lyulj@comp.nus.edu.sg
\IEEEcompsocthanksitem S. C.-K. Chau and N. Wang are with the Research School of Computer Science, Australian National University. \protect\\
E-mail: \{sid.chau, vincent.wang\}@anu.edu.au
\IEEEcompsocthanksitem Y. Zheng is with Data61, CSIRO, Marsfield NSW 2122, Australia. \protect\\
Email: yifeng.zheng@data61.csiro.au
\IEEEcompsocthanksitem The corresponding author is S. C.-K. Chau. 
\IEEEcompsocthanksitem This paper is to appear in IEEE Trans. Cloud Computing.
}
}


\IEEEtitleabstractindextext{
\begin{abstract}
Cloud computing has been a dominant paradigm for a variety of information processing platforms, particularly for enabling various popular applications of sharing economy. However, there is a major concern regarding data privacy on these cloud-based platforms. This work presents novel cloud-based privacy-preserving solutions to support collaborative consumption applications for sharing economy. In typical collaborative consumption, information processing platforms need to enable fair cost-sharing among multiple users for utilizing certain shared facilities and communal services. Our cloud-based privacy-preserving protocols, based on homomorphic Paillier cryptosystems, can ensure that the cloud-based operator can only obtain an aggregate schedule of all users in facility sharing, or a service schedule conforming to service provision rule in communal service sharing, but is unable to track the personal schedules or demands of individual users. More importantly, the participating users are still able to settle cost-sharing among themselves in a fair manner for the incurred costs, without knowing each other's private schedules or demands. Our privacy-preserving protocols involve no other third party who may compromise privacy. We also provide an extensive evaluation study and a proof-of-concept system prototype of our protocols.
\end{abstract}

}

\maketitle

\IEEEpeerreviewmaketitle

\begin{sloppypar}
\section{Introduction}

\IEEEPARstart{C}{loud} computing has been used as a dominant paradigm of computing systems for a variety of information processing platforms. A typical information processing platform requires users to submit their personal data, which will be processed and synthesized in centralized cloud-based systems. This paradigm underlies many e-commerce, social network and work productivity platforms. In particular, cloud-based platforms are used to support the emerging paradigm of ``sharing economy'', which aims to facilitate cost-sharing among end-users and promote efficient utilization of shared resources. Among the various applications in sharing economy, {\em collaborative consumption} is one of the major classes that enable the shared use of facilities or services by a group of users for reducing the cost of otherwise exclusive consumption. 

In collaborative consumption applications, there are certain shared facilities, resources and communal services that users can request and book in a dynamic fashion through some cloud-based platforms, which coordinate the utilization and cost-sharing among the involved users. For example, occupants can reserve personal resources (e.g., hot desks and dormitory beds) and share the cost of communal facilities and amenities (e.g., utilities, air conditioning, kitchens, bathrooms, gyms). Another example is about sharing communal services, such as transportation shuttles, garbage collection and postal services. These communal services are usually operated according to fixed schedules. To improve efficiency, these communal services should be operated according to user-initiated demands. Users should be able to submit their demands and schedule the services accordingly. At the end, the incurred cost will be shared among the participants who request the facilities or services. 

However, there is a heightened concern in data privacy nowadays. Consumer watchdog organizations have frequently voiced their concerns regarding misuse of user data without user consents. Notably, there were many well-known incidents of data breaches, whereby sensitive user data was stolen by hackers or leaked to public. Furthermore, more stringent data protection and privacy laws are being introduced worldwide. In particular, the General Data Protection Regulation (GDPR) was introduced in Europe, which requires strict data privacy protection measures on the cloud-based platforms that collect user data.

This work aims to address the privacy concern arising in cloud-based applications and systems. Especially, we present novel privacy-preserving solutions to enhance cloud-based information processing platforms for collaborative consumption applications in sharing economy. In this paper, we consider two important aspects on cloud-based platforms for collaborative consumption:
\begin{enumerate}

\item {\bf Privacy}: Typical users will be deterred from sharing the knowledge of personal usage, if this can compromise their privacy. We aim to ensure sufficient user privacy in the sharing activities without leaking personal data of users.

\item {\bf Fair Cost-Sharing}: The associated cost of collaborative consumption should be attributed fairly to the participants. Even users are unwilling to reveal the knowledge of personal usage, we still provide effective solutions to determine the fair cost-sharing contributions of each user.

\end{enumerate}

The goal of this work is to offer proper solutions that can enable fair cost-sharing and preserve privacy in collaborative consumption. Fortunately, the complete knowledge of personal usage is not always necessary for facility or service provision. Sometimes, only the coarse aggregate usage, rather than individual ones, suffices. Also, the detailed personal usage is not necessary for fair cost-sharing. Each user only needs to know the portion of his own usage to compute his contribution.

\subsection{Use Cases of Cloud-based Privacy-Preserving Collaborative Consumption}

This paper presents cloud-based privacy-preserving protocols for collaborative consumption. Through a {\em privacy-preserving} protocol, users can settle cost-sharing among themselves without knowing each others' usage, whereas the facility or service operators in the cloud are only revealed the minimal information of aggregate usage for facility or service provision. Furthermore, other than users and operators, there should be no other third party involved in the protocols, who may compromise privacy.

\begin{figure*}[!htp]
\centering
\includegraphics[width=0.95\textwidth]{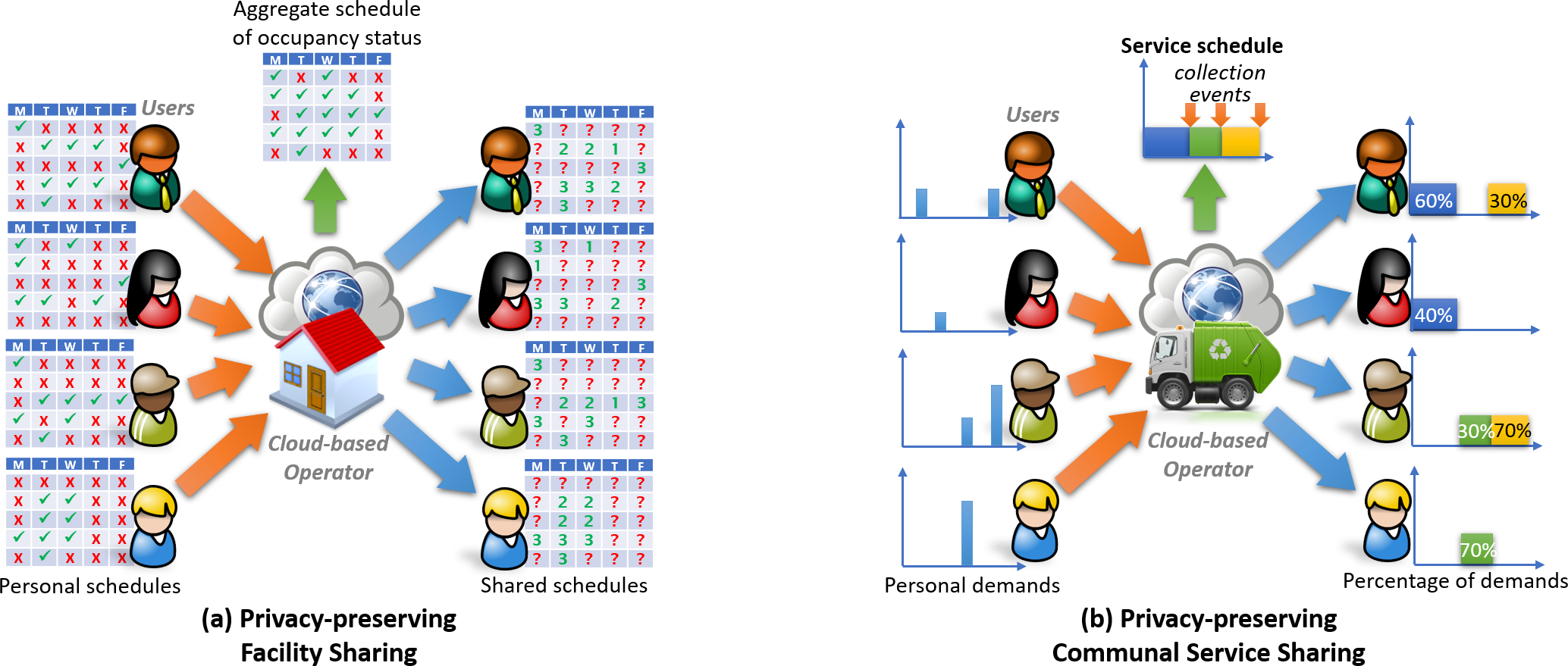}
\caption{Illustrations of (a) privacy-preserving facility sharing and (b) privacy-preserving communal service sharing.}
\label{fig:examples}
\end{figure*}

In particular, we consider two general use cases of cloud-based privacy-preserving collaborative consumption (as illustrated in Fig.~\ref{fig:examples}):
\begin{enumerate}

\item {\bf Privacy-Preserving Facility Sharing}: Users are supposed to share the cost of a certain facility in a privacy-preserving manner via a cloud-based platform that is managed by a facility operator. But the facility operator does not know the individual users' usage of the facility. More generally, even if the operator wants to optimize space allocation, he only needs to know the coarse occupancy (e.g., reserving a small room for 10 persons or less, and a big room for more than 10 persons). In Fig.~\ref{fig:examples} (a), users first submit their personal schedules of usage. With a proper privacy-preserving aggregation protocol, the operator can only obtain an aggregate schedule of binary (or coarse) occupancy status, but is unable to track the personal usage of individual users. Users will be provided only the number of concurrent users in their used timeslots, but not other timeslots that they do not use. This allows users to settle cost-sharing payment among themselves in a privacy-preserving manner. A time-varying access key to the facility will be provided to the requested users, after proper payment is received\footnote{Privacy-preserving payments can be handled using privacy-oriented  cryptocurrencies (e.g., Monero, Dash, ZCash).}. 

\item {\bf Privacy-Preserving Communal Service Sharing}: Users request their demands of a certain communal service in a timely manner via a cloud-based platform that is managed by a service operator. The service operator is supposed to provide the service when the accumulative demand exceeds a certain threshold based on a service provision rule. For example, in garbage collection service, a garbage collector is supposed to carry out a collection service to possibly multiple disposal sites, when the total accumulative amount of garbage disposal is sufficiently high. In Fig.~\ref{fig:examples} (b), users first submit their personal demands. With a proper privacy-preserving aggregation protocol, the operator can only obtain a service schedule conforming to the service provision rule of the accumulative demand exceeding a threshold, but is unable to track the personal demands of individual users. Users will be provided the percentage of their demands of each involved service action as a basis of service charge settlement among themselves.

\end{enumerate}

Although there are various cloud-based privacy-preserving protocols in other applications (e.g., ride-sharing), the novelty of our protocols is not only about enabling privacy-preserving data aggregation, but also privacy-preserving cost-sharing among users. Users will be provided information of their usage for only the respectively used timeslots to settle cost-sharing. Otherwise, users will be oblivious of the usage of others. This can mitigate privacy compromise. Also, privacy-preserving payment is incorporated in the applications of collaborative consumption. In this paper, our novel privacy-preserving solutions for collaborative consumption are based on homomorphic Paillier cryptosystems. 

This paper is organized as follows. We first review the related work in Section~\ref{sec:related}. We formulate the problems in Section~\ref{sec:model} and present the preliminary of cryptosystems in Section~\ref{sec:prelim}. The cloud-based privacy-preserving protocols are described in Section~\ref{sec:algo}, with a theoretic privacy analysis of our protocols in Section~\ref{sec:analysis}. The extensions of our protocols are provided in Section~\ref{sec:extension}. We present a proof-of-concept cloud-based prototype in Section~\ref{sec:sys} and an extensive empirical evaluation study of our protocols in Section~\ref{sec:eval}. Section~\ref{sec:concl} concludes this paper.

\section{Related Work} \label{sec:related}

\subsection{Cloud-based Data Aggregation and Privacy}
The rise of sharing economy has drawn substantial attention to proper design of its mechanisms. The previous studies mainly focus on the aspects of game-theoretical and big data analyses \cite{CE17sharing}. However, the privacy aspect on the related information processing platforms has been studied to a less extent. Recently, there are several studies about privacy-preserving ride-sharing \cite{ridesharingfortransferable,oride}. These studies aim to preserve the private locations of riders and drivers in matching. However, the extant privacy-preserving ride-sharing literature does not consider cost-sharing among riders and drivers, and hence does not address privacy-preserving cost-sharing.

As a departure from the prior literature, this work is the first study to address privacy-preserving collaborative consumption and cost-sharing on a cloud-based platform. In a broad sense, privacy-preserving collaborative consumption involves privacy-preserving data aggregation. Generally speaking, this line of work shares the common goal of learning an aggregation result from the data collected from a set of users without revealing individual users' private data. So far, a large body of work on privacy-preserving data aggregation (e.g., \cite{shi2011privacy,li2013efficient,erkin2013privacy,zheng2018learning,WCL20}) has been proposed in the literature. However, combining privacy-preserving data aggregation and privacy-preserving cost sharing has not been investigated in the prior work.

Another relevant work of cloud-based privacy-preserving data aggregation is privacy-preserving meeting scheduling~\cite{igor11privacymeeting}, where privacy-preserving protocols were proposed by taking advantage of the homomorphic properties of well-known cryptosystems to privately and efficiently compute common availabilities among users. However, their meeting scheduling problem is different from our problems, and they did not consider cost-sharing. 

To summarize, the previous work has not addressed cloud-based privacy-preserving collaborative consumption. Our study is the first work to integrate cloud-based privacy-preserving data aggregation and privacy-preserving cost-sharing. Also, the prior work of cloud-based privacy-preserving data aggregation mostly considers exact data aggregation, whereas our protocols also support {\em coarse data aggregation} (e.g., coarse occupancy) to mask the true aggregate data by an estimation function.

\subsection{Privacy-Preserving Techniques and Systems}
We also briefly review privacy-preserving techniques, such as differential privacy, secure multi-party computation and homomorphic encryption.

\begin{enumerate}

\item
{\em Differential privacy}: Dwork et.al.~\cite{dwork2006calibrating} first proposed the notion of \emph{differential privacy} (DP) for the single database scenario, where a trusted database server \emph{answers queries} in a privacy-preserving manner with tailored randomization to the response to the query~\cite{dwork2006calibrating}. In the absence of a trusted aggregator (e.g., cloud applications), users need to perturb their data before submission to the aggregator. Without the combination of cryptographic techniques and distributed data randomization, each user has to add large enough noise to guarantee local differential privacy~\cite{shi2011privacy}. Compared with cryptographic techniques, differential privacy always comes with a trade-off between privacy and accuracy, hence not suitable for privacy-preserving cost-sharing. 

\item
{\em Secure multi-party computation (MPC)}: The problem of computing a multi-party sum where no party reveals its value in the clear, even to the aggregator, is referred to as \emph{Secure Aggregation}~\cite{bonawitz2017practical}. However, most MPC constructions are interactive. Therefore, directly employing MPC in our setting would require users to interact with each other whenever a coarse aggregate statistic needs to be computed. Such an interaction model may not be desirable in practical settings, especially in a client-server computation model as often seen in cloud computing applications~\cite{yao2015using,wu2019secedmo}.

\item
{\em Homomorphic encryption}: Fully homomorphic cryptosystems can support general computation on ciphertexts but less efficient which result in a solution that is impractical~\cite{gentry2009fully}. On the other hand, partially homomorphic encryption is less costly, such as Paillier~\cite{paillier1999public}. Most previous work on homomorphic encryption considers homomorphic operations on ciphertexts encrypted under the same key~\cite{boneh2005evaluating,gentry2009fully}. These schemes do not directly apply in our problems, since if users encrypted their data under the aggregator's public key, the aggregator would not only be able to decrypt the aggregate statistics, but also users' private data. Therefore, similar to \cite{igor11privacymeeting}, we rely on homomorphic properties of asymmetric cryptosystems to prevent the operator from gaining any knowledge about the privacy of individual users.

\end{enumerate}

In terms of privacy-preserving system architecture, there are distributed or centralized designs. Distributed design relies on equal computation from its participants. However, distributed design requires frequent and intensive message exchanges among participants, presenting limitations in scalability and complexity for a large number of (resource-limited) mobile devices. Two additional drawbacks are the need of sequencing among participants and the unpredictability of results (e.g., if a participant interrupts the protocol).
On the other hand, centralized cloud-based design \cite{igor11privacymeeting, qin2018privacy} is better in terms of scalability, communication cost, and resilience. Therefore, we draw on the centralized design but also carefully address the privacy issues using homomorphic encryption.

\section{Problem Formulation} \label{sec:model}

This section formulates the general problems of privacy-preserving collaborative consumption.

\subsection{General Problem Setup} 

There are two parties in our problems: (1) an operator managing a cloud-based platform, and (2) a group of $N$ users, denoted by $i \in \{1 . . . N\}$, who are sharing a certain resource or service via the cloud-based platform. 

In general, there are some common properties in the privacy-preserving literature:
\begin{enumerate}

\item {\em Honest-but-curious Model}: We assume the honest-but-curious (or semi-honest) model, where the operator and users always obey the protocol instructions by disclosing truthful information, but they may attempt to explicitly or implicitly infer the private information of others based on the intermediate results obtained during the execution of the protocol. This model follows a practical setting, because the operator always has incentive for business with customers and the users have the genuine needs for resources or services from the operator. It is the primary interest of everyone to allow the business proceeds as intended, and there are little incentives for them to behave maliciously. However, the personal data may be leaked intentionally or unintentionally from the operator or a user to a third party (e.g., data selling, or data breach). We also assume that all users have no-zero usage or demand.

\item {\em No Collusion}:
We assume that there is no collusion among these semi-honest users
. This is a realistic assumption, because collusion will involve disclosing personal information to others, which will defeat the purpose of privacy.

\item {\em Privacy with Cloud-based Operator}:
The cloud-based operator is expected to be given the {\em minimal} information for their operations and services, such as a coarse aggregate schedule of usage, or a service schedule confirming the service provision rule. However, the personal data of users, such as personal usage, should not be revealed or inferred from the protocol by the operator. 

\item {\em Privacy with Other Users}:
Each user is expected to be given the portion of his own usage in the incurred resources or services. This allows a user to determine his contribution of cost-sharing. However, the personal data of other users, such as personal usage, should not be revealed or inferred from the protocol by the user. 

\end{enumerate}

Next, we formally define three collaborative consumption problems: uncapacitated facility sharing (UFS), capacitated facility sharing (CFS), and communal service sharing (CSS). For clarity, we first provide the problems' definitions without considering privacy-preserving protocols. In Section~\ref{sec:algo}, we will provide privacy-preserving protocols to solve these problems using homomorphic cryptosystem (to be explained in Section~\ref{sec:prelim}). Also, Table~\ref{tbl:symbs} lists key symbols and notations in the problem formulations.

\begin{table}[!t]
    \centering\caption{\label{tbl:symbs}Key Symbols and Notations.}
    \scalebox{0.9}{\begin{tabularx}{\linewidth}{c|X}
    \hline    \hline
    Symbol & Definition \\
    \hline
   UFS &  Uncapacitated Facility Sharing problem \\	
   CFS &  Capacitated Facility Sharing problem \\
   CSS &  Communal Service Sharing problem \\
   $N$  & Total number of users \\
   $i$  & The $i$-th user  \\
   $m$  & Total number of timeslots \\
   $b_{i}^{j}$  & Requested facility usage of user $i$ at the $j$-th timeslot in UFS/CFS\\
   $N^j$ & Total number of requested users at the $j$-th timeslot in UFS/CFS\\   
   $p_{i}^{j}$ & Requested demand of user $i$ at the $j$-th timeslot  in CSS\\
   $q_{i}^{k}$ & Fraction of requested demand of user $i$ in the $k$-th service action in CSS\\
   $s^k$ & Timeslot of the $k$-th service action in CSS\\   
   
    \hline    \hline
    \end{tabularx}}
\end{table}

\subsection{Uncapacitated Facility Sharing (UFS)}

We first define the Uncapacitated Facility Sharing (UFS) problem, which is illustrated in Fig.~\ref{fig:examples} (a). In this problem, the users share a certain facility, and the operator only needs to know whether the facility will be used by any users at a specific timeslot (coarse aggregate), without considering the actual capacity, i.e., total number of requested users at the timeslot (exact aggregate). 
In UFS, each user $i$ has a private usage schedule of using a certain facility over discrete timeslots, denoted by a sequence of bits $b_i = [b_{i}^{1}, ..., b_{i}^{m}]$. Each bit $b_{i}^{j} \in \{0, 1\}$ represents the usage request of user $i$ at the $j$-th timeslot, where $b_{i}^{j} = 1$ means that user $i$ requests usage of the facility at the $j$-th timeslot, whereas $b_{i}^{j} = 0$ means that the user does not request usage at the timeslot. Note that the number of timeslots $m$ is constant for all users. 

The users are supposed to provide the operator with $(b_i)_{i=1}^N$ in a privacy-preserving manner (without revealing the individual schedule $b_i$ of each $i$). In return, each user $i$ is given $a_i = [a_{i}^{1}, ..., a_{i}^{m}]$, where 
\begin{equation}
a_{i}^{j}  \triangleq  \left\{ \begin{array}{cl}
N_j, & \mbox{if \ } b_{i}^{j} = 1 \\
?, & \mbox{if \ } b_{i}^{j} = 0
\end{array}\right.
\end{equation}
where $N_j = \sum_{i=1}^N b_{i}^{j}$ is the number of requested users at the $j$-th timeslot. Namely, if $i$ requests usage at the $j$-th timeslot, then $i$ is given the total number of requested users at the timeslot. Otherwise, $i$ is unable to infer the information of the number of requested users at the $j$-th timeslot.  The users who request usage will split the associated cost based on the number of requested users at each timeslot.

The operator will only be able to derive the coarse aggregate schedule of usage $c = [c^{1}, ..., c^{m}]$, where $c^{j} = 1$ means that there exists at least user $i$ with $b_{i}^{j} = 1$, and $c^{j} = 0$ otherwise. Namely,
\begin{equation}
c^j \triangleq \bigvee_{i=1}^N b_{i}^{j}
\end{equation}
where $\vee$ is the binary disjunction operator.

\subsection{Capacitated Facility Sharing (CFS)}

We next define the Capacitated Facility Sharing (CFS) problem, where the operator also needs to estimate the number of requested users of each timeslot coarsely for the purpose of capacity allocation. For example, a building operator will reserve a small room if the number of users is less than 10, or a larger room otherwise.
Consider a sequence $(C_1, ..., C_r)$ such that $0 < C_1 < ... < C_r \le N$. Each $C_r$ represents of the capacity of the $r$-th facility. For convenience, we let $C_0 = 0$. We define an estimation function ${\sf f}(\cdot) \in \{0, ..., r \}$ by:
\begin{equation}
{\sf f}(x) = r, \mbox{\ if\ } C_{r -1} < x \le C_{r}
\end{equation}
Estimation function ${\sf f}(\cdot)$ maps the total number of users $x$ to the required facility $r$ that can provide the minimal sufficient capacity. ${\sf f}(\cdot)$ can mask $x$, without revealing the true number of users.

In CFS, the users are supposed to provide the operator with $(b_i)_{i=1}^N$ in a privacy-preserving manner. In return, each user $i$ is given $a_i$ (same as that of UFS). But the operator will only be able to derive a schedule of required facilities $\tilde{c} = [\tilde{c}^{1}, ..., \tilde{c}^{m}]$, where 
\begin{equation}
\tilde{c}^j \triangleq {\sf f}\Big( \sum_{i=1}^N b_{i}^{j} \Big) = {\sf f}(N^j).
\end{equation}
Note that UFS is a sub-problem of CFS by setting $C_1 = N$.

{\bf Remarks:}  UFS and CFS can capture different degrees of privacy with cloud-based operators. In UFS, the operator does not know the number of occupants in each timeslot, but only the status of being occupied or not, whereas CFS allows the operator know only coarse occupant number, specified by the thresholds of capacities. We note that CFS can allow arbitrary levels of coarse occupant number, which makes our protocol flexible to capture any degree of privacy requirement with the operator.

\subsection{Communal Service Sharing (CSS)}

Finally, we define the Communal Service Sharing (CSS) problem, which is illustrated in Fig.~\ref{fig:examples} (b). In this problem, each user has a schedule of personal demands of a service over discrete timeslots, denoted by a sequence of real numbers $p_i = [p_{i}^{1}, ..., p_{i}^{m}]$, where $p_{i}^{j}$ represents the requested demand of user $i$ at the $j$-th timeslot. 
The users are supposed to provide the operator with $(p_i)_{i=1}^N$ in a privacy-preserving manner. 
The operator will be able to derive a schedule of service actions. Each service action is determined by a simple service provision rule -- if the accumulative demand from all users since the last service action exceeds a threshold $C$, then a new service action is required. Formally, we denote the schedule of service actions by $s = [s^{1}, ..., s^{k}, ...]$, where $s^k$ is the timeslot of the $k$-th service action that is defined by:
\begin{equation}
s^k \triangleq \min\Big\{ \ t \ \mid \ t \ge s^{k-1}+1 \mbox{ \ and \ }  \sum_{i=1}^N \sum_{j = s^{k-1}+1}^{t}    p_{i}^{j} \ge C \Big\}.
\end{equation}
For example, $p_i$ represents the amount of garbage disposal from user $i$ over time. A garbage service operator will dispatch a garbage collection service when the total amount of garbage disposal from all users is sufficiently large by exceeding $C$.

In return, each user $i$ is given $q_i = [q_{i}^{1}, ..., q_{i}^{k}, ...]$, where 
\begin{equation}
q_{i}^{k} \triangleq 
\left\{ \begin{array}{cl}
\frac{\sum_{j = s^{k-1}+1}^{s^{k}}  p_{i}^{j} } { \sum_{i'=1}^{N}  \sum_{j = s^{k-1}+1}^{s^{k}} p_{i'}^{j}},  & \mbox{\ if\ } \sum_{j = s^{k-1}+1}^{s^{k}}  p_{i}^{j}  > 0 \\
0, & \mbox{ otherwise \ } 
\end{array}\right.
\end{equation}
Namely, if $i$ has non-zero demand in the $k$-th service action, then $i$ is given $q_{i}^{k}$, the fraction of his demand in that service action. The users will split the associated cost based on the portions of their demands in all service actions.

\medskip

{\bf Remarks:} All UFS, CFS and CSS problems share similar structures of usage data aggregation at the operator and cost-sharing among users. UFS is the most basic one. CFS extends UFS by considering facility capacity, whereas CSS extends UFS by considering service action scheduling.

\section{Preliminary of Cryptosystems} \label{sec:prelim}

Our privacy-preserving solutions rely on several standard tools from cryptosystems (e.g., Paillier cryptosystem). We briefly explain these tools in this section. More detailed explanations can be found in standard cyber security and cryptosystem textbooks (e.g., \cite{YiPB14}).

\subsection{Asymmetric Key Cryptosystems}

Many cryptosystems use some forms of asymmetric key cryptosystems. We assume that the users share a common secret, which is used to derive a common key pair $(K_P , K_s)$, where $K_p$ is the public key and $K_s$ is the private key. This can be accomplished, for example, through a secure credential establishment protocol (\cite{C08key,L09key}).
The private key is derived and known to each user but not to the operator. We write the encryption of a message $x$ with the group public key by ${\mathbb E}_{K_P, r}[x] = y$, where $r$ is a randomly chosen integer (that does not affect the decryption),  and the decryption of the encrypted message $y$ by ${\mathbb D}_{K_s}[y] = x$.

\subsection{Homomorphic and Paillier Cryptosystems}

Homomorphic cryptosystems allow computational operations on encrypted data, with results that match those of the operations as if they had been performed on unencrypted data (plaintext). Homomorphic cryptosystems are often used for privacy-preserving computation. This allows data to be encrypted and out-sourced for third-party computation (e.g., cloud-based platforms) without revealing the true data. 

Paillier cryptosystem \cite{YiPB14} is a realization of partially homomorphic cryptosystem, which satisfies the following homomorphic properties:
\begin{equation}
    {\mathbb D}_{K_s}\Big[{\mathbb E}_{K_p, r_1}[m_1] * {\mathbb E}_{K_p, r_2}[m_2] \  (\mbox{mod\ }n^{2}) \Big] = m_1+m_2 \ (\mbox{mod\ } n) \label{eqn:paillier-add}
\end{equation}
\begin{equation}
   {\mathbb D}_{K_s}\Big[({\mathbb E}_{K_p, r_1}[m_1])^{m_2}  \  (\mbox{mod\ }n^{2})\Big] = m_1 \cdot m_2   \ (\mbox{mod\ } n) \label{eqn:paillier-mul}
\end{equation}
$K_p$ and $K_s$ are respectively the public and private keys. $n$ is generated by $n=p \cdot q$ where $p$ and $q$ are random large prime numbers. $m$ is the plaintext to be encrypted. $r_1,r_2 \in \mathbb{N}$ are any randomly chosen positive integers, which will not affect the outcomes of decryption. For brevity, we will suppress the  (\mbox{mod}) terms in the equations of this paper.


\subsection{Paillier Cryptosystem with Negative and Non-Integer Numbers}

In the original Paillier cryptosystem, the value space is $\mathbb{Z}_{\sf N} = \{0, 1, \cdots, {\sf N} - 1\}$. To handle negative numbers and real numbers, we follow the approach in \cite{wang2012harnessing}. Since the homomorphic property of Paillier cryptosystem is over modulo arithmetic, we can shift the negative numbers to the interval of $\{\frac{{\sf N} - 1}{2} + 1, ..., {\sf N} - 1\}$, such that we set $-1 = {\sf N} -1$ mod ${\sf N}$, while keeping non-negative numbers in the interval of $\{0, 1, ... , \frac{{\sf N} - 1}{2}\}$. When the decrypted value $y_i > \frac{{\sf N}}{2}$, we shift $y_i$ back by $y_i - {\sf N}$. This can limit the input values to the interval of $(-\frac{{\sf N}}{2}, \frac{{\sf N} - 1}{2}]$. We set ${\sf N}$ to have 1024 bits to be sufficiently large for accommodating all the computations in our protocols.

For real number $y \in \mathbb{R}^n$ (or $y_i \in \mathbb{R}$), we can use Scaling, Rounding, Unscaling (SRU) mechanism \cite{wang2012harnessing,miao2015cloud}. Let $S$ be the scaling factor, where $S$ should be large enough to eliminate rounding errors. Then we can approximate the value of $y_i \in \mathbb{R}$ by $\bar{y}_i = \lfloor S \cdot y_i \rfloor \in \mathbb{Z}_{\sf N}$. Here we use $\bar{y}_i$ to denote the rounded off integer of $y_i$. We will use similar notations for other variables in this paper. The approximate value of $y_i$ can be recovered by dividing $S$ (i.e., $\frac{y_i}{S}$). {\sf N}ote that this may result in an inaccuracy problem. To ensure acceptable rounding errors, the scaling factor should be properly chosen considering the specific application context. 

\subsection{Threshold Paillier Cryptosystem}

To improve the key distribution in Paillier cryptosystem, it is possible that the users do not need to share a common private key. A variant of Paillier cryptosystem is called $(N, t)$-threshold Paillier cryptosystem, in which the private key $K_s$ is divided and distributed to $N$ users (denoted as $K_{s1}, K_{s2}, \cdots, K_{sN}$) during the initialization stage, such that each user only obtains a share of the private key, rather than the complete private key. To decrypt a ciphertext, a user needs to cooperate with at least $t-1$ other users. Each user $i$ ($1 \leq i \leq N$) computes a partial decryption $c_i =c^{2N!K_{si}}$ of ciphertext $c = {\mathbb E}_{K_p} (m)$ with his own partial private key $K_{si}$. At least $t$ copies of $c_i$ will be required to decrypt $c$. An $(N, t)$-threshold Paillier cryptosystem has been proposed in \cite{damgaard2001generalisation}.

\medskip

\section{Cloud-based Privacy-Preserving Protocols} \label{sec:algo}

In this section, we present cloud-based privacy-preserving protocols to solve UFS, CFS and CSS problems using Paillier cryptosystem. A theoretic privacy analysis of our protocols is given in Section~\ref{sec:analysis}.

\subsection{Privacy-Preserving Uncapacitated Facility Sharing (PP-UFS)} \label{sec:PPUFS}

First, we present a privacy-preserving protocol to ensure privacy in uncapacitated facility sharing. Our protocol is called Privacy-Preserving Uncapacitated Facility Sharing (PP-UFS) protocol, which is consisted of four stages as follows:

\subsubsection{Stage 0: Initialization}
\

In the initialization stage, the users derive a common key pair $(K_P , K_s)$, through a secure credential establishment protocol. The common key pair $(K_P , K_s)$ will be used in Paillier cryptosystem. Each user $i$ also generates a random number $r_i^j$ for each timeslot $j$ for encryption in Paillier cryptosystem.

In the protocol, most encryption and decryption operations  are carried out by the users. But the operator also needs to encrypt a random integer $R^j$ for each timeslot $j$, without revealing it to the users. Thus, the operator is given the public key $K_P$ -- but not the private key $K_s$, so cannot decrypt any encrypted messages.

\subsubsection{Stage 1: Distribution to Users}
\

\begin{figure}[t]
\centering 
\includegraphics[scale=0.5]{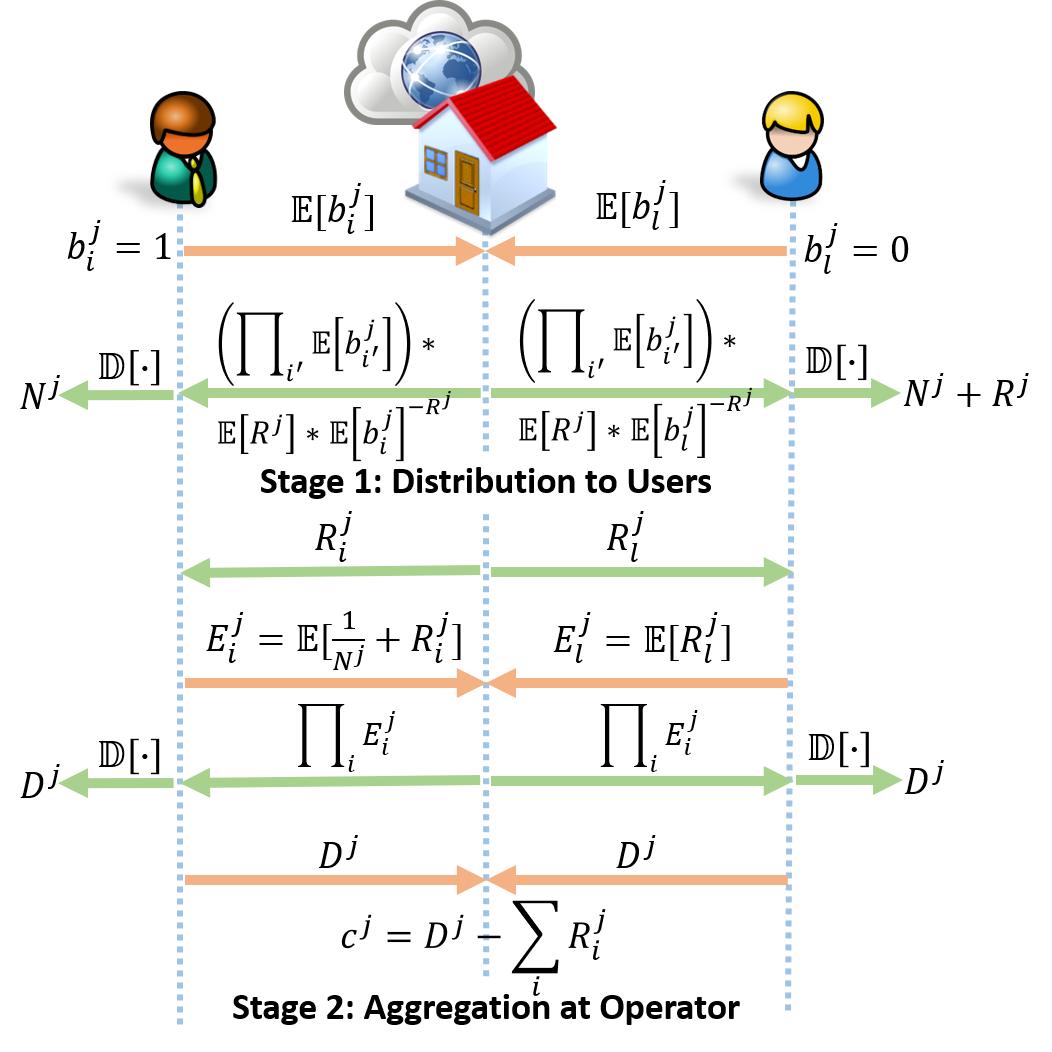} 
\caption{An illustration of PP-UFS Stages 1 and 2.}
\label{fig:protocol}
\end{figure}

In this stage, the users provides their personal schedules $(b_i)_{i=1}^N$ in a privacy-preserving manner. If user $i$ requests usage at the $j$-th timeslot (i.e., $b_{i}^{j} = 1$), then $i$ is given $N^j$ (the number of requested users at the timeslot).  See Fig.~\ref{fig:protocol} for an illustration of this stage. 

The steps of this stage are executed as follows:

\begin{enumerate}

\item Each user $i$ first encrypts $b_{i}^{j}$ and sends ${\mathbb E}_{K_p, r_i^j}[b_{i}^{j}]$ to the operator for each timeslot $j$.

\item
The operator multiplies all the received encrypted value $({\mathbb E}_{K_p, r_i^j}[b_{i}^{j}])_{i=1}^N$ as $\prod_{i'=1}^N {\mathbb E}_{K_p, r_{i'}^j}[b_{i'}^{j}]$ with an encrypted random integer ${\mathbb E}_{K_p, r}[R^j]$ (obtained in Stage 0). The operator also takes each ${\mathbb E}_{K_p, r_i^j}[b_{i}^{j}]$ to the power of $-R^j$, and returns the following product to each user $i$ for each timeslot $j$:
\begin{equation}
\big( \prod_{i'=1}^N {\mathbb E}_{K_p, r_{i'}^j}[b_{i'}^{j}] \big) * {\mathbb E}_{K_p, r}[R^j] * {\mathbb E}_{K_p, r_i^j}[b_{i}^{j}]^{-R^j}
\end{equation}

\item Each user $i$ receives and decrypts the following value for each timeslot $j$: 
\begin{align*}
& {\mathbb D}_{K_s}\Big[\big( \prod_{i'=1}^N {\mathbb E}_{K_p, r_{i'}^j}[b_{i'}^{j}] \big) *{\mathbb E}_{K_p, r}[R^j] * {\mathbb E}_{K_p, r_i^j}[b_{i}^{j}]^{-R^j} \Big]\\
= & \ {\mathbb D}_{K_s}\Big[ {\mathbb E}_{K_p, r_i^j}[\sum_{i=1}^N b_{i}^{j}+R^j -R^j \cdot b_{i}^{j}] \Big] \\
= & \ \left\{ \begin{array}{ll}
{\mathbb D}_{K_s}\Big[{\mathbb E}_{K_p, r_i^j}[ \sum_{i=1}^N b_{i}^{j}] \Big], & \mbox{if \ } b_{i}^{j} = 1 \\
{\mathbb D}_{K_s}\Big[{\mathbb E}_{K_p, r_i^j}[ \sum_{i=1}^N b_{i}^{j}+R^j] \Big], & \mbox{if \ } b_{i}^{j} = 0
\end{array}\right.
\end{align*}
which follows the homomorphic addition property (Eqn.~\raf{eqn:paillier-add}) and the homomorphic multiplication property (Eqn.~\raf{eqn:paillier-mul}) of Paillier cryptosystem.
If $b_{i}^{j} = 1$, then $i$ decrypts as follows:
\begin{equation}
{\mathbb D}_{K_s}\Big[{\mathbb E}_{K_p, r_i^j}[ \sum_{i=1}^N b_{i}^{j}] \Big] =  \sum_{i=1}^N b_{i}^{j} = N^j
\end{equation} 
Otherwise, if $b_{i}^{j} = 0$, then $i$ can only obtain $N^j +R^j$, which is not useful because of the unknown random integer $R^j$.

\end{enumerate}

\subsubsection{Stage 2: Aggregation at Operator}
\

In this stage, the operator derives the coarse aggregate schedule of usage $c = [c^{1}, ..., c^{m}]$, where $c^{j} = 1$ means there exists at least one user $i$ with $b_{i}^{j} = 1$, and $c^{j} = 0$ otherwise. See Fig.~\ref{fig:protocol} for an illustration of this stage.
The steps of this stage are executed as follows:

\begin{enumerate}

\item 
The operator first generates a random integer $R_i^j$ for each user $i$ and timeslot $j$. The operator sends $(R_i^j)_{j=1}^m$ to user $i$. 

\item
Recall $N^j \triangleq \sum_{i=1}^N b_{i}^{j}$. If $b_{i}^{j} = 1$, then $i$ knows $N^j$ from Stage 1. Hence, $i$ encrypts and sends the following value to the operator for each timeslot $j$:
\begin{align}
E_i^j \triangleq \ \left\{ \begin{array}{ll}
{\mathbb E}_{K_p, r_i^j}\Big[\frac{1}{N^j} + R_i^j\Big], & \mbox{if \ } b_{i}^{j} = 1 \\
{\mathbb E}_{K_p, r_i^j}\Big[R_i^j\Big], & \mbox{if \ } b_{i}^{j} = 0 \\
\end{array}\right.
\end{align}

\item
The operator returns $\prod_{i=1}^N E_i^j$ to all users for each timeslot $j$.

\item
Each user $i$ receives and decrypts the same value ($\prod_{i=1}^N {E}_i^j$) as $D^j \triangleq {\mathbb D}_{K_s}\Big[\prod_{i=1}^N E_i^j \Big]$. There are two cases:

\medskip
\begin{enumerate}

\item[(i)] If $b_{i}^{j} = 1$ for at least one user $i$, then $D^j = 1 + \sum_{i=1}^N R_i^j$.

\item[(ii)] If $b_{i}^{j} = 0$ for every user $i$, then $D^j = \sum_{i=1}^N R_i^j$.

\end{enumerate}

\medskip
This follows the homomorphic addition property of Paillier cryptosystem (Eqn.~\raf{eqn:paillier-add}).
Note that the users with $b_{i}^{j} = 0$ cannot determine which of the above cases is true, because of the unknown random sum $\sum_{i=1}^N R_i^j$.

\item 
All users return the same decrypted value $D^j$ to the operator. Because the operator knows $\sum_{i=1}^N R_i^j$, he can determine the following cases:

\medskip
\begin{enumerate}

\item[(i)]  If $D^j - \sum_{i=1}^N R_i^j = 1$, then the $j$-th timeslot is used by at least one user.

\item[(ii)]  If $D^j - \sum_{i=1}^N R_i^j = 0$, then the $j$-th timeslot is used by none of users.

\end{enumerate}

\medskip
Hence, the operator sets $c^j = D^j - \sum_{i=1}^N R_i^j$.

\end{enumerate}

\subsubsection{Stage 3: Cost-sharing and Payment}
\

In the final stage, the users will pay the usage fee to the operator. The operator should publish the usage fee rate for each timeslot in advance. Each user who uses a particular timeslot has known $N^j$, the number of requested users at the timeslot (from Stage~1). They will pay the usage fee proportionally to $\frac{1}{N^j}$. Note that payments to the operator can be attained in a privacy-preserving manner by privacy-oriented cryptocurrencies (e.g., Monero, Dash, ZCash). 

The operator will not grant facility access to the users, unless the required amount of usage fee has been fully collected. Since the users are semi-honest, they will pay the associated fees in order to gain the access. Note that the facility access can be implemented as a form of secret keys that can unlock the facility (e.g., using smart locks)\footnote{Note that the secret keys do not need to be physically entered by the users. The secret keys can be downloaded to smartphones or RFID cards, which can communicate automatically and wirelessly to the smart locks.}. 

The secret keys can be distributed to the users  in a privacy-preserving manner as follows. Suppose a secret key is represented by $\kappa^j$. The operator sends ${\mathbb E}_{K_p, r_i^j}[b_{i}^{j}]^{\kappa^j}$ to each user $i$. Hence, only the users with $b_{i}^{j} = 1$ (that has been submitted in Stage 1) can decrypt the secret key for facility access: 
\begin{equation}
{\mathbb D}_{K_s} \big[ {\mathbb E}_{K_p, r_i^j}[b_{i}^{j}]^{\kappa^j}  \big] = {\mathbb D}_{K_s} \big[ {\mathbb E}_{K_p, r_i^j}[b_{i}^{j} \cdot {\kappa^j}]  \big] = \kappa^j.
\end{equation}

\medskip

\subsection{Privacy-Preserving Capacitated Facility Sharing  (PP-CFS)} \label{sec:PPCFS}

Next, we present a privacy-preserving protocol for capacitated facility sharing, called Privacy-Preserving Capacitated Facility Sharing (PP-CFS) protocol, which is a generalization of PP-UFS. PP-CFS is consisted of four stages as follows:

\subsubsection{Stage 0: Initialization}
\

This stage is the same as Stage 0 in PP-UFS.

\subsubsection{Stage 1: Distribution to Users}
\

This stage is the same as Stage 1 in PP-UFS.


\subsubsection{Stage 2: Aggregation at Operator}
\

This stage is similar to Stage 2 in PP-UFS with slight modifications.
In this stage, the operator derives a schedule of required facilities $\tilde{c} = [\tilde{c}^{1}, ..., \tilde{c}^{m}]$, where $\tilde{c}^j = {\sf f}(N^j)$.

The steps of this stage are executed as follows:

\begin{enumerate}

\item 
The operator first generates a random integer $R_i^j$ for each user $i$ and timeslot $j$. 
The operator sends $(R_i^j)_{j=1}^m$ to user $i$. 

\item
Each user $i$ encrypts and sends the following value to the operator for each timeslot $j$:
\begin{align}
\hat{E}_i^j \triangleq \ \left\{ \begin{array}{ll}
{\mathbb E}_{K_p, r_i^j}\Big[\frac{{\sf f}(N^j)}{N^j} + R_i^j\Big], & \mbox{if \ } b_{i}^{j} = 1 \\
{\mathbb E}_{K_p, r_i^j}\Big[R_i^j\Big], & \mbox{if \ } b_{i}^{j} = 0 \\
\end{array}\right.
\end{align}

\item
The operator returns $\prod_{i=1}^N \hat{E}_i^j$ to all users for each timeslot $j$.

\item
Each user $i$ receives and decrypts the same value as $\hat{D}^j \triangleq {\mathbb D}_{K_s}\Big[\prod_{i=1}^N \hat{E}_i^j \Big]$. 

There are two cases of $\hat{D}^j$:

\medskip
\begin{enumerate}

\item[(i)] If $b_{i}^{j} = 1$ for at least one user $i$, then $\hat{D}^j = {\sf f}(N^j) + \sum_{i=1}^N R_i^j$.

\item[(ii)] If $b_{i}^{j} = 0$ for every user $i$, then $\hat{D}^j = \sum_{i=1}^N R_i^j$.

\end{enumerate}

This follows the homomorphic addition property of Paillier cryptosystem (Eqn.~\raf{eqn:paillier-add}).

Note that the users with $b_{i}^{j} = 0$ still cannot determine the value of ${\sf f}(N^j)$, due to the unknown random sum $\sum_{i=1}^N R_i^j$.

\item 
All users return the same decrypted value $\hat{D}^j$ to the operator. Because the operator knows $\sum_{i=1}^N R_i^j$, he can determine the following cases:

\medskip
\begin{enumerate}

\item[(i)]  If $\hat{D}^j - \sum_{i=1}^N R_i^j = {\sf f}(N^j)$, then the $j$-th timeslot is used by at least one user.

\item[(ii)]  If $\hat{D}^j - \sum_{i=1}^N R_i^j = 0$, then the $j$-th timeslot is used by none of users.

\end{enumerate}

\medskip
Hence, the operator sets $\tilde{c}^{j} = \hat{D}^j - \sum_{i=1}^N R_i^j$.

\end{enumerate}

\subsubsection{Stage 3: Cost-sharing and Payment}
\

This stage is similar to Stage 3 in PP-UFS. However, the operator sets the usage fee rate proportional to ${\sf f}(N^j)$. The users requesting usage at the respective timeslot have known $N^j$, and then can determine the proper usage fee rate and the corresponding payments.

\medskip

\subsection{Privacy-Preserving Communal Service Sharing (PP-CSS)} \label{sec:PPCSS}

Lastly, we present a privacy-preserving protocol for communal service sharing, called Privacy-Preserving Communal Service Sharing (PP-CSS) protocol with four stages as follows:

\subsubsection{Stage 0: Initialization}
\

This stage is the same as Stage 0 in PP-UFS.

\subsubsection{Stage 1: Aggregation at Operator}
\

Unlike PP-UFS, in the first stage, the operator derives a schedule of service actions $s = [s^{1}, ..., s^{k}, ...]$, where $s^k$ is the timeslot of the $k$-th service action when the accumulative demand from all users since the last service action at timeslot $s^{k-1}$ exceeds threshold $C$. 

Suppose $s^{k-1}$ is known for some $k$. Otherwise, we start with $s^{0} = 0$. The steps of this stage are executed as follows:

\begin{enumerate}

\item[(0)] Let $l = s^{k-1}$ and $t = l + 1$.

\item
Each user $i$ encrypts and sends the following value to the operator:
\begin{equation}
E_i^{(l,t)}  \triangleq  {\mathbb E}_{K_p, r_i^k}\Big[\sum_{t'=l+1}^t p_{i}^{t'}  - \frac{C}{N} \Big]
\end{equation}

\item
The operator generates a random positive number $R^t$ and returns the product $\prod_{i=1}^N (E_i^{(l,t)})^{R^t}$ to all users.

\item
Each user $i$ receives and decrypts the same value ($\prod_{i=1}^N (E_i^{(l,t)})^{R^t}$) as:
\begin{equation}
D^t \triangleq {\mathbb D}_{K_s}\Big[\prod_{i=1}^N (E_i^{(l,t)})^{R^t} \Big] = R^t \cdot (\sum_{i=1}^N \sum_{t'=l+1}^t p_{i}^{t'} - C)
\end{equation}
which follows the homomorphic addition property of Paillier cryptosystem  (Eqn.~\raf{eqn:paillier-add}). Note that no user can determine the value of $\sum_{i=1}^N \sum_{t'=l+1}^t p_{i}^{t'}$ because of unknown random number $R^t$. However, they can determine if $D^t$ is positive or negative. All users send the indicator variable ${\mathds 1}(D^t \ge 0)$ to the operator.

\item
The operator receives ${\mathds 1}(D^t \ge 0)$. There are two cases:
\medskip
\begin{enumerate}

\item[(i)] If $D^t \ge 0 $, then set the timeslot of the next service action by $s^k = t$. Next, set $k \leftarrow k + 1$, go to Step (0) and repeat until $t > m$.

\item[(ii)] If $D^t < 0 $, then set $t \leftarrow t + 1$,  go to Step (1) and repeat until $t > m$.

\end{enumerate}

\item The operator obtains a schedule of service actions $[s^{1}, ..., s^{k}, ...]$.

\end{enumerate}

\subsubsection{Stage 2: Distribution to Users}
\

Let the total demand of user $i$ in the $k$-the service action be $P_i^k \triangleq \sum_{j = s^{k-1}+1}^{s^{k}}  p_i^j$. In the second stage, each user $i$ derives $q_i = [q_{i}^{1}, ..., q_{i}^{k}, ...]$, where $q_{i}^{k} = 
\frac{P_i^k} {\sum_{i'=1}^{N} P_{i'}^k}$. This stage is a generalization of Stage 1 in PP-UFS, because $P_i^k$ can be fractional, instead of binary.

\begin{enumerate}

\item Each user $i$ first encrypts $P_i^k$ and sends ${\mathbb E}_{K_p, r_i^k}[P_i^k]$ to the operator for the $k$-th service action.

\item The operator multiplies all the received encrypted value $({\mathbb E}_{K_p, r_{i'}^k}[P_{i'}^k])_{{i'}=1}^N$ as $\prod_{{i'}=1}^N {\mathbb E}_{K_p, r_{i'}^k}[P_{i'}^k]$ with an encrypted random integer ${\mathbb E}_{K_p, r}[R^k]$ (obtained in Stage 0), and returns the following product to each user $i$ for each service action~$k$:
\begin{equation}
\big( \prod_{i'=1}^N {\mathbb E}_{K_p, r_{i'}^k}[P_{i'}^k] \big) * {\mathbb E}_{K_p, r}[R^k] 
\end{equation}

\item Each user $i$ computes and sends the following product to the operator for each service action $k$:
\begin{equation}
\Big( \big( \prod_{i'=1}^N {\mathbb E}_{K_p, r_{i'}^k}[P_{i'}^k] \big) * {\mathbb E}_{K_p, r}[R^k] \Big)^{P_i^k} *{\mathbb E}_{K_p, r_i^k}[0]
\end{equation}
Note that the operator cannot infer $P_i^k$ because ${\mathbb E}_{K_p, r_i^k}[0]$ is encrypted using an unknown random number $r_i^k$ and the operator does not have the private key.

\item The operator returns the following product to each user $i$  for each service action $k$:
\begin{align}
& 
\Big( \big( \prod_{i'=1}^N {\mathbb E}_{K_p, r_{i'}^k}[P_{i'}^k] \big) * {\mathbb E}_{K_p, r}[R^k] \Big)^{P_i^k} *  \nonumber \\
&{\mathbb E}_{K_p, r_i^k}[0] *{\mathbb E}_{K_p, r_i^k}[P_i^k]^{-R^k}
\end{align}

\item Each user $i$ receives and decrypts the following value for each service action $k$: 
\begin{align}
& {\mathbb D}_{K_s}\Big[\Big( \big( \prod_{i'=1}^N {\mathbb E}_{K_p, r_{i'}^k}[P_{i'}^k] \big) * {\mathbb E}_{K_p, r}[R^k] \Big)^{P_i^k} *  \nonumber \\
&\qquad \quad {\mathbb E}_{K_p, r_i^k}[0] *{\mathbb E}_{K_p, r_i^k}[P_i^k]^{-R^k} \Big]  \\
= & \ {\mathbb D}_{K_s}\Big[ {\mathbb E}_{K_p, r_i^k}[P_i^k \cdot (\sum_{i'=1}^N P_{i'}^k+R^k)] *  \nonumber \\
&\qquad \quad {\mathbb E}_{K_p, r_i^k}[0]*{\mathbb E}_{K_p, r_i^k}[-R^k \cdot P_i^k] \Big] \\
= &\ \left\{\begin{array}{ll}
{\mathbb D}_{K_s}\Big[{\mathbb E}_{K_p, r_i^k}[ P_i^k \cdot(\sum_{i'=1}^N P_{i'}^k)] \Big]\\
\quad \  \ =P_i^k \cdot \sum_{i'=1}^N P_{i'}^k, & \mbox{if \ } P_i^k > 0 \\
{\mathbb D}_{K_s}\Big[{\mathbb E}_{K_p, r_i^k}[ 0 ] \Big]=0, & \mbox{if \ } P_i^k=0
\end{array}\right.
\end{align}

which follows the homomorphic addition property (Eqn.~\raf{eqn:paillier-add}) and the homomorphic multiplication property (Eqn.~\raf{eqn:paillier-mul}) of Paillier cryptosystem.
If $P_i^k > 0$, then $i$ divides $P_i^k \cdot \sum_{i'=1}^N P_{i'}^k$ by $P_i^k$ to obtain $\sum_{i'=1}^N P_{i'}^k$ and $q_{i}^{k} = 
\frac{P_i^k} {\sum_{i'=1}^{N} P_{i'}^k}$.

\end{enumerate}

\subsubsection{Stage 3: Cost-sharing and Payment}
\

This stage is similar to Stage 3 in PP-UFS. But each user $i$ pays a service fee that is proportional to $q_i^k$ for the $k$-th service action.  Each user $i$ requesting demand in the $k$-th service action has been aware of $q_i^k$ and can determine his proportion of service fee accordingly.

\medskip

\section{Security Analysis} \label{sec:analysis}

In this section, we provide a security analysis of PP-UFS, PP-CFS and PP-CSS. We show that all protocols can ensure privacy with the operator and other users.

\begin{theorem}
Given the semi-honest and no-collusion adversary assumption and the privacy of the homomorphic cryptosystem, our PP-UFS protocol ensures that (i) the operator learns nothing except whether the facility will be used by any users at a timeslot $j$, and (ii) a user $i$ learns nothing other than the number of requested users at the timeslot $j$ when $i$ has $b_i=1$ or learns nothing when $b_i=0$.
\end{theorem}

\textbf{Security analysis of PP-UFS}.
We are going to split our analysis into two cases: the case where a user say $i$ is corrupted and the case where the operator is corrupted. Intuitively, we want to show a corrupted party cannot learn any unexpected information from its view, i.e., the transcript of messages obtained from the protocol execution \cite{HazayL10}. Note that as the third stage of the protocol is simply based on the secure computation result from the first two stages for settling payment, our security analysis just needs to be focused on the first two stage.

We first analyze the privacy against a corrupted user $i$.
In Stage 1, the view of a user $i$ is $\sum_{i=1}^N b_{i}^{j}+R^j-R^j \cdot b_{i}^{j}$. If $i$ has $b^j_i=1$, then $i$ only obtains the aggregate usage $\sum_{i=1}^N b_{i}^{j}$ at timeslot $j$, which is allowed by definition. Otherwise, $i$ only obtains a randomly masked aggregate schedule, i.e., $\sum_{i=1}^N b_{i}^{j}+R^j$, which is indistinguishable from random values, so $i$ learns no information. In Stage 2, the view of a user $i$ is $R^j_i$ and $1 + \sum_{i=1}^N R_i^j$ or $\sum_{i=1}^N R_i^j$. These are all values which are indistinguishable from random values, so each user $i$ learns no information from the view. Based on the composition theorem \cite{canetti2000security}, we can have the privacy guarantees that a user $i$ learns no more than the number of requested users at the timeslot $j$ when $i$ has $b_i=1$ or learns nothing when $b_i=0$.

We now analyze the privacy against the operator. In Stage 1, the view of the operator is ${\mathbb E}_{K_p,r^j_i}[b_{i}^{j}]$. In Stage 2, the view of the operator is $E_i^j$, and $1 + \sum_{i=1}^N R_i^j$ or $\sum_{i=1}^N R_i^j$, equivalent to $1$ (resp. $1 + \sum_{i=1}^N R_i^j$) or $0$ (resp. $\sum_{i=1}^N R_i^j$) as the operator knows $\sum_{i=1}^N R_i^j$. Therefore, except the allowed-by-definition result about whether a facility will be used by any users at timeslot $i$, the operator only observes entirely ciphertexts which are protected by semantically secure homomorphic cryptosystem. The privacy of homomorphic cryptosystem ensures that the operator learns no information from these ciphertexts. Based on the composition theorem \cite{canetti2000security}, we can have the privacy guarantees that the operator learns nothing except whether the facility will be used by any users at timeslot $j$.

\begin{theorem}
Given the semi-honest and no-collusion adversary assumption and the privacy of the homomorphic cryptosystem, our PP-CFS scheme ensures that (i) the operator learns nothing except a coarse estimation of the capacity at a timeslot $j$, and (ii) a user $i$ learns nothing other than the number of requested users at the timeslot $j$ when $i$ has $b_i=1$ or learns nothing when $b_i=0$.
\end{theorem}

\textbf{Security analysis of PP-CFS}. The security analysis is almost the same as the analysis for the PP-UFS protocol, so we omit the details here. Note that the PP-CFS is just a generalization of the PP-UFS protocol, with only one difference in that in Stage 2 of PP-CFS, each user $i$ encrypts $\frac{{\sf f}(N^j)}{N^j}$ as opposed to $\frac{1}{N^j}$ in PP-UFS, so as to allow the operator to obtain a coarse estimation of the capacity, rather than a binary aggregate schedule in PP-UFS.

\begin{theorem}
Given the semi-honest and no-collusion adversary assumption and the privacy of the homomorphic cryptosystem, our PP-CSS scheme provides the following guarantees: (i) the operator learns no information other than whether a service action should be made; (ii) all users can learn whether the accumulated demand exceeds the threshold without knowing. Meanwhile, a user with zero demand learns no more information, while a user with non-zero demand can learn the exact accumulated demand for computing service fees.
\end{theorem}

\textbf{Security analysis of PP-CSS}. Similar to the analysis for the PP-UFS scheme, we analyze the view of a user $i$ and the view of the operator respectively. The view of a user $i$ in Stage 1 is $R^t \cdot (\sum_{i=1}^N \sum_{t'=l+1}^t p_{i}^{t'} - C)$, from which a user only learns if the accumulative demand from all users since the last service action exceeds a threshold or not, but not the exact accumulative demand due to the randomness $R^t$ introduced by the operator. The view of a user $i$ in Stage 2 is $\sum_{i'=1}^N P_{i'}^k+R^k$, $P_{i}^k\cdot \sum_{i'=1}^N P_{i'}^k=P_{i}^k(\sum_{i'=1}^N P_{i'}^k+R^k)-P_{i}^k\cdot R^k$. So, if $i$ has $P^k_i > 0$, the view becomes $\sum_{i'=1}^N P_{i'}^k+R^k$ and $P_{i}^k\cdot \sum_{i'=1}^N P_{i'}^k$, so $i$ can obtain $\sum_{i'=1}^N P_{i'}^k$ (through $(P_i^k \cdot \sum_{i'=1}^N P_{i'}^k) / P_i^k$) and hence $q_i^k$, which is as expected. If $i$ has $P^k_i = 0$, the view becomes $\sum_{i'=1}^N P_{i'}^k+R^k$ and $0$. Here, $\sum_{i'=1}^N P_{i'}^k+R^k$ is a value indistinguishable from random value, from which $i$ gains no information. So $i$ with $P^k_i = 0$ learns nothing from the view of Stage 2. Based on the composition theorem, the privacy guarantee against the user follows.

The view of the operator in Stage 1 is $E_i^{(l,t)}$ and ${\mathds 1}(D^t \ge 0)$. The indicator variable ${\mathds 1}(D^t \ge 0)$ is expected to known by the operator so that a service action can be determined. For the ciphertext $E_i^{(l,t)}$, the operator learns no information due to the semantic privacy of homomorphic cryptosystem. Therefore, the operator learns no information other than whether a service action needs to be scheduled. The view of operator in Stage 2 is ${\mathbb E}_{K_p, r_i^k}[P_i^k]$ and ${\mathbb E}_{K_p, r_i^k}[P_i^k \cdot (\sum_{i'=1}^N P_{i'}^k+R^k)]*{\mathbb E}_{K_p, r_i^k}[0]$. Clearly, these are entirely ciphertexts and the operator learns no information. Note that here ${\mathbb E}_{K_p, r_i^k}[0]$ is to prevent the case that a user has $P_i^k=1$ and thus the ciphertext ${\mathbb E}_{K_p, r_i^k}[P_i^k \cdot (\sum_{i'=1}^N P_{i'}^k+R^k)]$ returned to the operator remains unchanged, making the operator know that $P_i^k=1$. Based on the composition theorem, we thus know that the operator learns nothing except whether a service action should be made.

\section{Extensions of Protocols} \label{sec:extension}

\subsection{Threshold Paillier Cryptosystem}

In this section, we extend our protocols such that the users do not need to share a common private key in Paillier cryptosystem. We rely on $(N, t)$-threshold Paillier cryptosystem~\cite{damgaard2001generalisation,cramer2001multiparty}. 

In the $(N, t)$-threshold Paillier cryptosystem, the additive homomorphic property is still preserved, while the private key is shared among $N$ parties. The decryption of a ciphertext requires a certain threshold number $t$ of parties to produce decryption shares with their decryption key share, and these decryption shares can be combined to produce the plaintext. We explain how user $i$ can recover the plaintext value underlying the ciphertext sent by the operator as follows:

\begin{enumerate}

\item 
A set of threshold decryption keys are generated and distributed by a trusted third party or generated online by a specific protocol;

\item 
The operator randomly selects $t-1$ users, and sends the ciphertext $C$ to them for partial decryption;

\item 
Each selected user $k$ produces and sends its decryption share $C_k$ to the operator;

\item 
The operator sends the ciphertext $C$ along with the $t-1$ decryption shares to the user $i$;

\item 
User $i$ computes a decryption share using his private key share. Then, user $i$ combines all decryption shares to produce the plaintext value.

\end{enumerate}

\subsection{User Misrepresentation Avoidance}
We also discuss how to address the issue of user misrepresentation, whereby the users inject falsified information in the protocols for malicious purposes, such as disrupting the execution of the protocol or gaining private information about other users. But note that the attack by user misrepresentation is not able to succeed in revealing the availabilities of other users, as the best case, a malicious user can set its own schedule to all-available, and then guesses the availabilities of the other $N-1$ users. 

For example, in PP-UFS in Section~\ref{sec:PPUFS}, a malicious user may modify his schedule during the execution of the protocol by changing the value of $b_i^j$ in different stages, however, this user is anyway unable to gain the sum of the availabilities of the other $N-1$ users if his $b_i^j=0$ in Stage 1 (1), because the randomness $R^j$ obtained in Stage 0 cannot be canceled during the decryption in Stage 1 (3). Thus, our privacy-preserving protocols are robust against user misrepresentation.

Our focus is on assuring data confidentiality against the operator. Malicious users are not our major consideration. But more complex techniques such as zero-knowledge proof (ZKP) can be integrated into our protocols to resist the potential malicious users, which will be a subject of future work.

\section{Cloud-based Prototype} \label{sec:sys}

\begin{figure}[!htp]
\centering
\begin{subfigure}[b]{0.32\textwidth}
\includegraphics[scale=0.29]{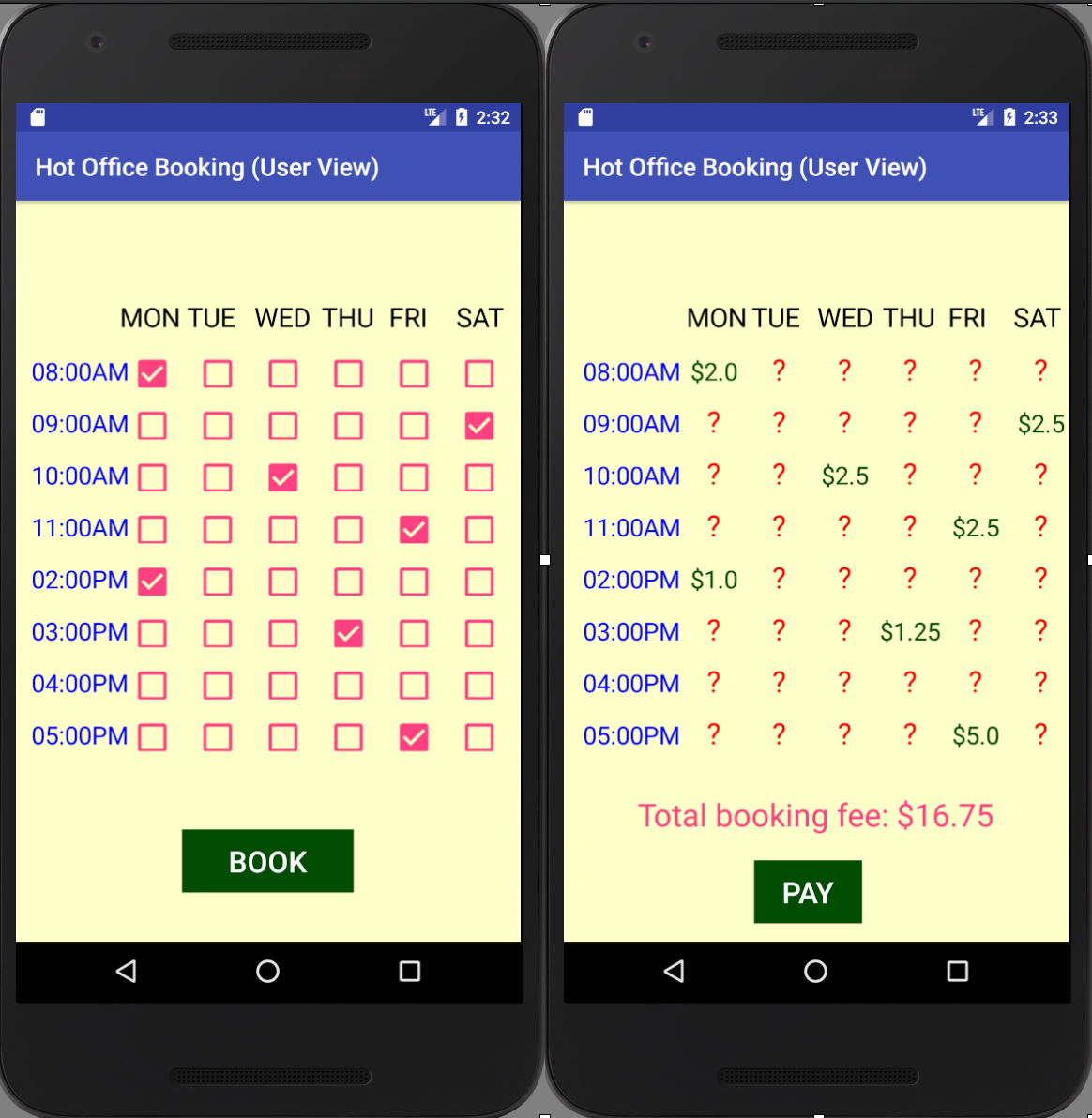}
\caption{User View.}
\label{fig:user}
\end{subfigure}
\begin{subfigure}[b]{0.16\textwidth}
\includegraphics[scale=0.29]{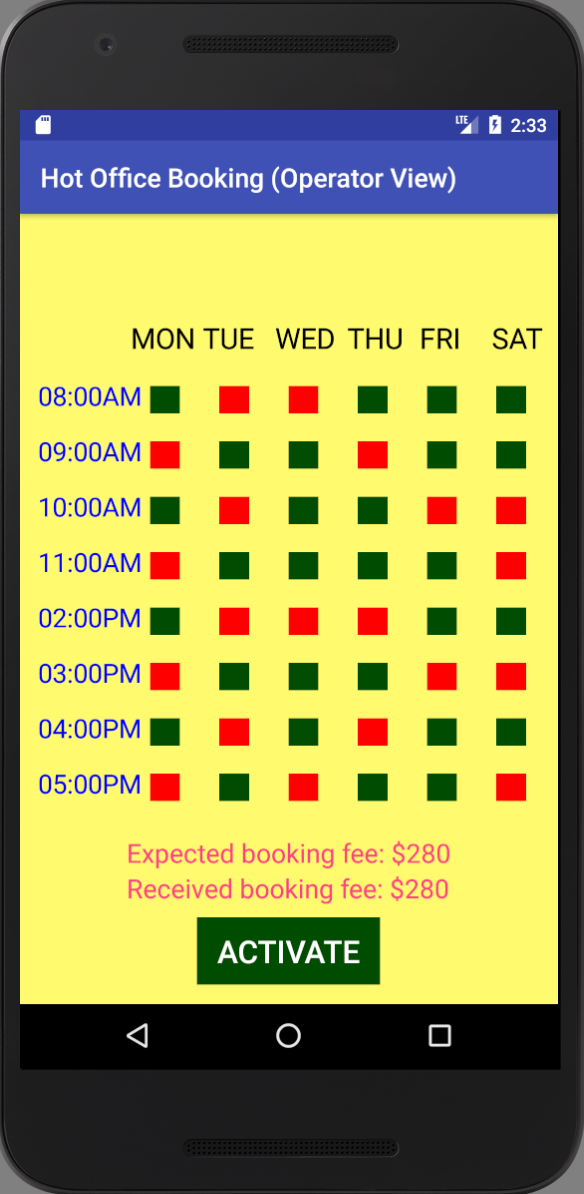}
\caption{Operator View.}
\label{fig:operator}
\end{subfigure}
\caption{Hot office booking cloud-based mobile app system.}
\end{figure}

We implemented the privacy-preserving protocols in a proof-of-concept cloud-based prototype for hot office booking. We used the \cite{paillier} for the implementation of homomorphic Paillier cryptosystems. The cloud computing server-side was implemented in Java 1.8.0 and the client-side was implemented on Android 7.0 smartphones. The cloud-based mobile app user interfaces are depicted in Fig.~\ref{fig:user}.

In this system, users are able to set up their bookings for a hot office between 8AM and 5PM from Monday to Saturday in a week. Users' preferences will be submitted through our privacy-preserving protocols to the hot office operator without compromising any  users' private schedules. The operator will perform computations on the encrypted data and each user will be provided with the total number of users of each timeslot that they have booked. Users' view mobile app interfaces show their partial payments for the hot office at their booked timeslots or ``?'' for non-booked timeslots. Meanwhile, the interfaces also show the total booking fee that a user needs to pay for these bookings. The operator's view mobile app interface is depicted in Fig.~\ref{fig:operator}, where green tiles and red tiles respectively indicate the occupied and vacant timeslots of the hot office. From the interface, the operator is able to visualize the distribution of binary occupancy status in the hot office as well as the total expected booking fee and the booking fee that has been received. Once the expected payments are received, the operator can confirm the bookings by pressing ``ACTIVATE'' button in order to distribute the access keys to users for their booked timeslots.

\section{Empirical Evaluation} \label{sec:eval}
\subsection{Experimental Setup}
For each user in PP-UFS, we randomly generate 48 values of \{0,1\} to represent 48 timeslots over 6 days, with each day having 8 timeslots. For each user in each timeslot in PP-CSS, we randomly generate real values from [10,20] and set the service threshold $C=100$. 
To evaluate the accuracy and sensitivity of our privacy-preserving protocols with respect to different parameters, we use \emph{Mean Relative Error} (MRE) to measure the relative distance between the decrypted aggregate schedule and the true aggregate schedule in PP-UFS, and the decrypted aggregate demand of service and the true aggregate demand of service in PP-CSS. We explicitly define MRE by averaging over all the $m$ timeslots as follows:
\begin{equation}
\text{MRE}=\frac{{\textstyle\sum}_{t=1}^m}{m}\frac{\abs{{\mathbb D}\Big[{\mathbb E} \big[{\textstyle\sum}_{i=1}^N x_i ^t\big]\Big]-{\textstyle\sum}_{i=1}^N x_i ^t}}{{\textstyle\sum}_{i=1}^N x_i ^t}
\end{equation}
where $x_i ^t$ represents the service schedule or service demand of user $i$ at timeslot $j$.

\begin{figure*}[ht]
\centering
\begin{subfigure}[b]{0.45\textwidth}
\includegraphics[width=0.9\linewidth]{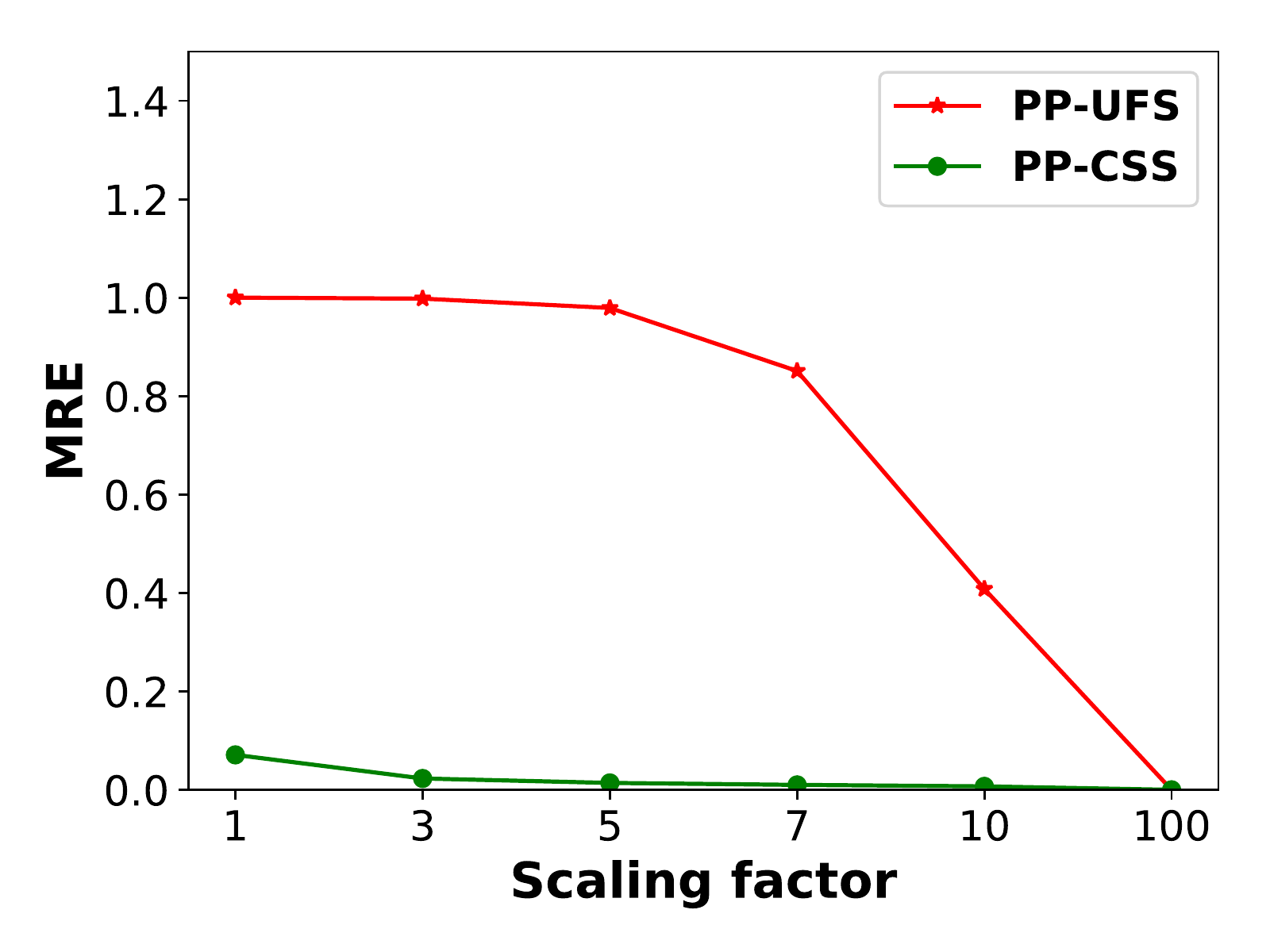}
\caption{MRE vs. Scaling factor.}
\label{MRE_scaling}
\end{subfigure} \qquad \qquad 
\begin{subfigure}[b]{0.45\textwidth}
\includegraphics[width=0.9\linewidth]{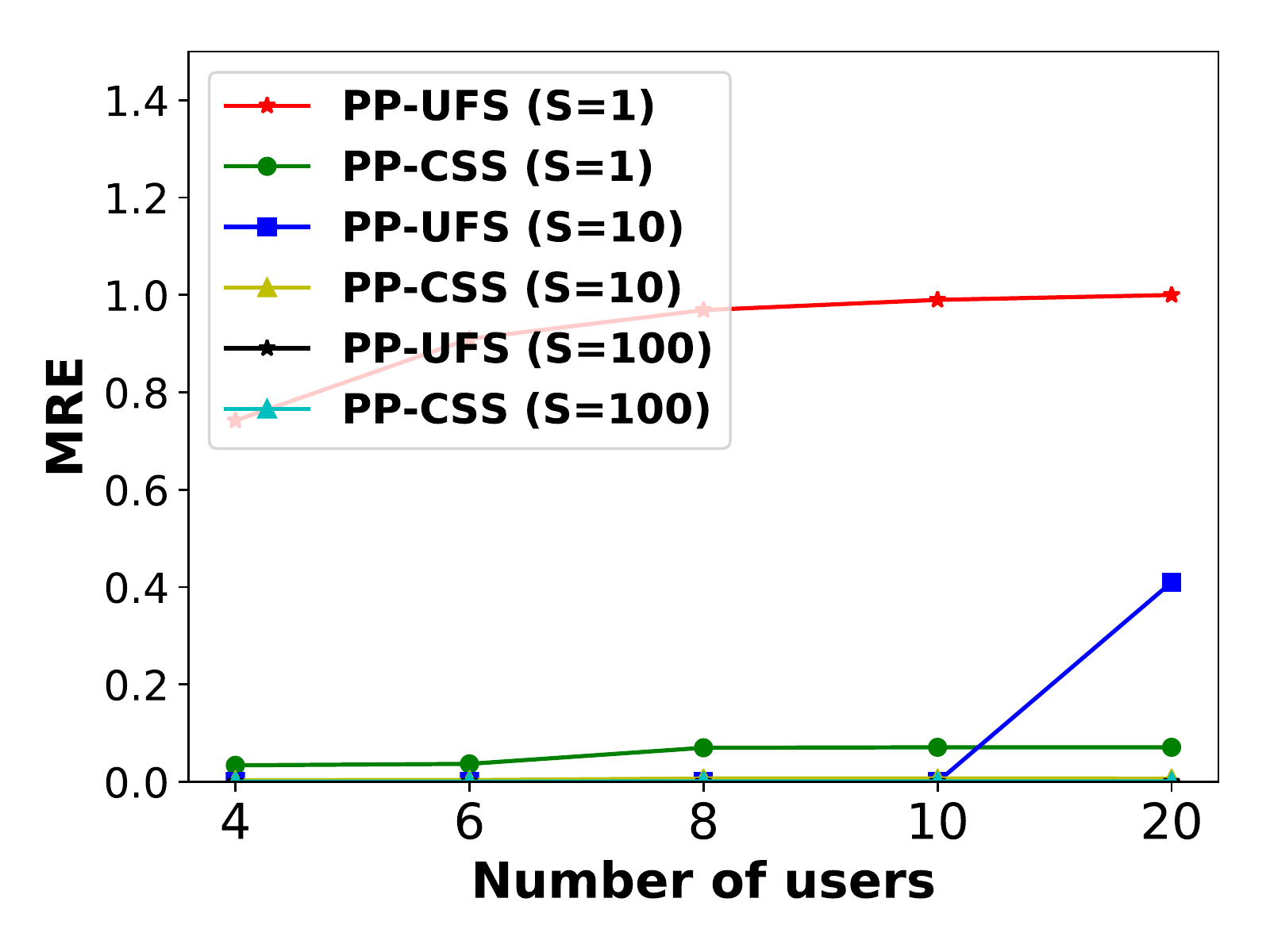}
\caption{MRE vs. Number of users.}
\label{MRE_users}
\end{subfigure}
\caption{MRE with varying scaling factor $S$ and number of users $N$ in PP-UFS and PP-CSS.}
\label{fig:MRE}
\end{figure*}

\subsection{Experimental Results}
\subsubsection{Accuracy and Sensitivity}
For privacy-preserving facility sharing, since the encryption and decryption operations in PP-UFS and PP-CFS are similar, here we only analyze PP-UFS for simplicity. As indicated in Section~\ref{sec:PPUFS}, the only potential MRE in PP-UFS comes from Stage2 (2) by scaling $1/N^j$, where $N^j$ is the total number of requested users at timeslot $j$, hence positively proportional to the number of users. By contrast, in PP-CSS, since the requested demand of user $i$ at timeslot $j$ can be fractional instead of binary, each encryption operation requires scaling. Moreover, as indicated in Section~\ref{sec:PPCSS}, Stage1 (1) needs to scale $C/N$, where $N$ is the total number of users, hence also positively proportional to the number of users. Below, we investigate how MRE varies with scaling factor $S$ when the number of users is fixed, and how MRE varies with the number of users when the scaling factor $S$ is fixed.

1. {\em MRE vs. Scaling Factor $S$}: 
Scaling aims to minimize the effect of rounding on utility. If $S$ is chosen to be $10^N$, which means considering $N$ digits to the right of the decimal point; If $S=1$, then the right of the decimal point is neglected, and the scalar is rounded to the nearest integer. Since MRE is only introduced by the scaling factor $S$, we vary the values of $S$ and measure the corresponding MRE. The experiment is repeated for 20 times, and we report the average results. Fig.~\ref{MRE_scaling} studies the impact of scaling factor $S$ on MRE by fixing the number of users as 20. For both PP-UFS and PP-CSS, this figure shows that as $S$ becomes larger, MRE starts to gradually degrade as expected. In particular, MRE becomes 0 when the scaling factor $S \geq 10$ for PP-CSS and $S \geq 100$ for PP-UFS, which implies a higher value of $S$ can provide negligible utility loss. 

2. {\em MRE vs. Number of Users $N$}:
We further investigate the impact of the number of users on MRE by varying the number of users ranging from 4 to 20 under the fixed scaling factor $S \in \{1,10,100\}$. As shown in Fig.~\ref{MRE_users}, when $S=1$, MRE is relatively large across different number of users for both PP-UFS and PP-CSS, agreeing with the large MRE with $S=1$ reported in Fig.~\ref{MRE_scaling}. More importantly, when $S=1$, MRE generally increases with the number of users, which would significantly compromise the utility. The reason behind this gradual increment in MRE is rather obvious: with the increasing number of users, $N^j$ becomes larger and $1/N^j$ becomes smaller, scaling factor $S=1$ rounds $1/N^j$ to the nearest integer would significantly reduce even negate the precision. In contrast, for PP-CSS with $S \geq 10$, the values of MRE are all reduced to 0, while PP-UFS requires $S \geq 100$ to ensure the negligible values of MRE, agreeing with $S \geq 100$ for PP-UFS reported in Fig.~\ref{MRE_scaling}. Therefore, a larger value of scaling factor is always preferred so as to minimise the effect of rounding on MRE in larger number of users.

\subsubsection{Computational Cost}
\
 
\begin{figure}[!t]
\centering
\begin{subfigure}[b]{0.23\textwidth}
\includegraphics[width=\linewidth]{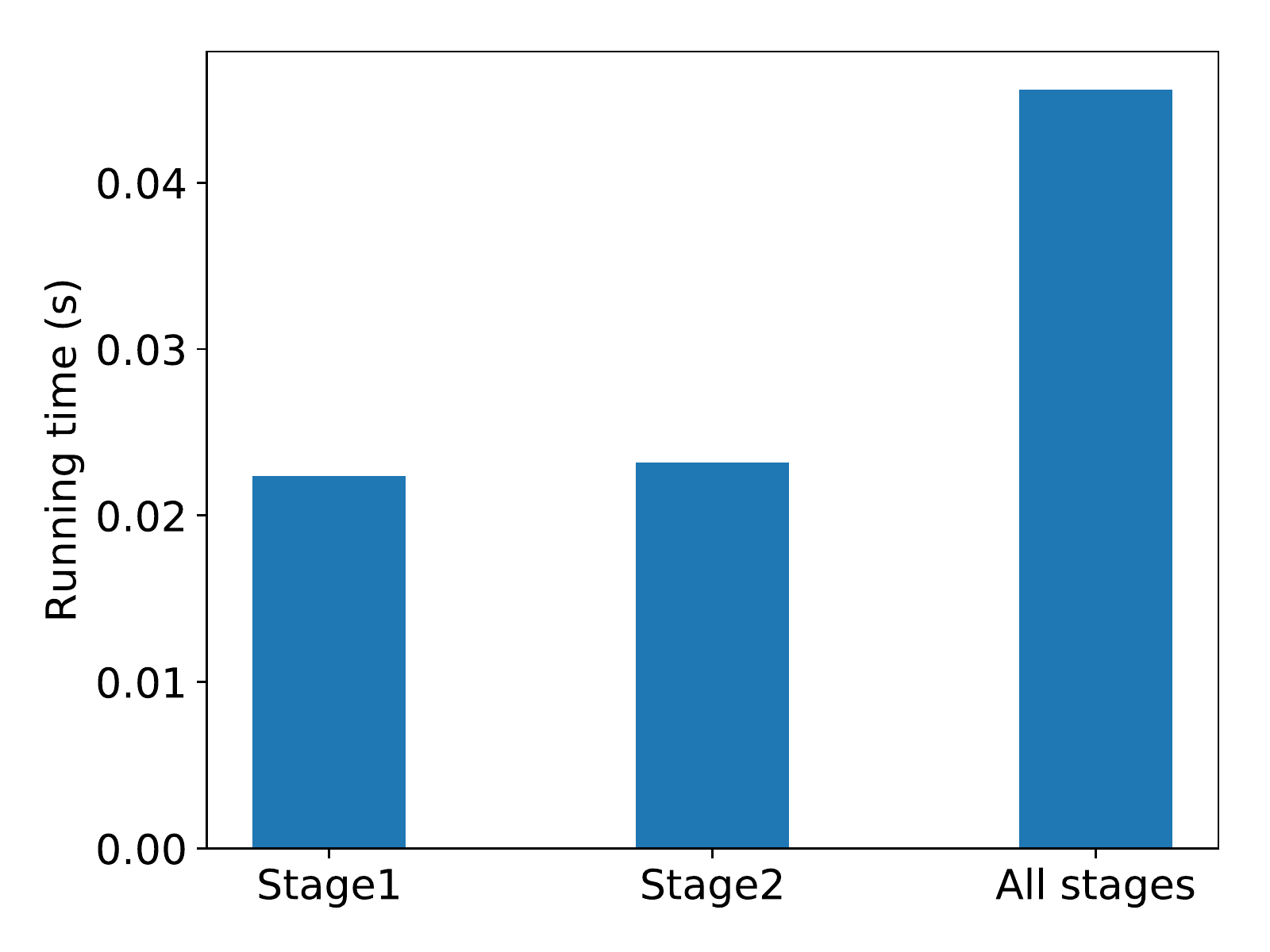}
\caption{}
\label{PPUFS_user_time}
\end{subfigure}\quad 
\begin{subfigure}[b]{0.23\textwidth}
\includegraphics[width=\linewidth]{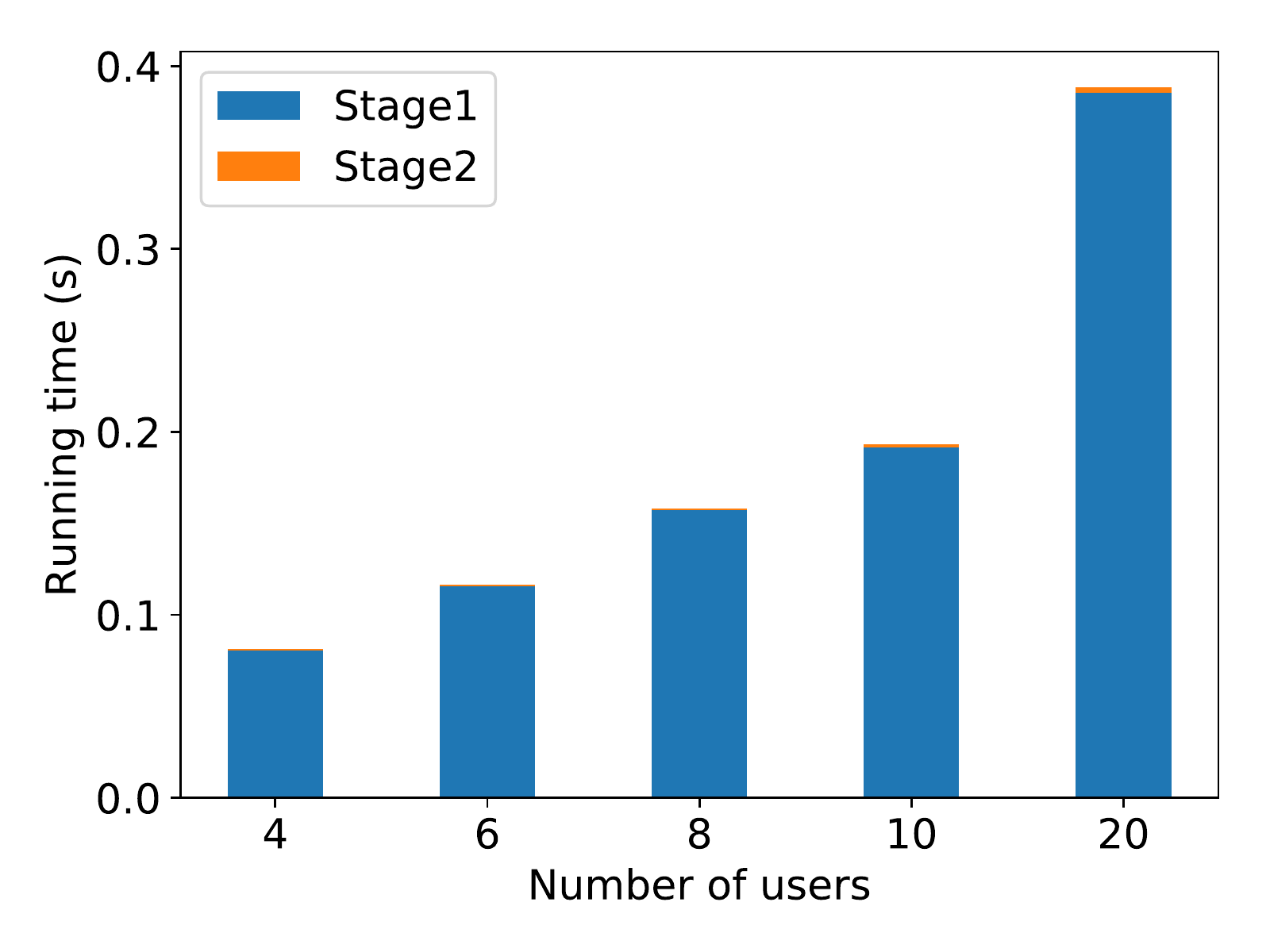}
\caption{}
\label{PPUFS_operator_time}
\end{subfigure}
\caption{PP-UFS: (a) Average user running time of different stages. (b) Operator running time vs. number of users.}
\label{fig:PPUFS_time}  
\end{figure}

\begin{figure}[!t]
 \centering
\begin{subfigure}[b]{0.23\textwidth}
\includegraphics[width=\linewidth]{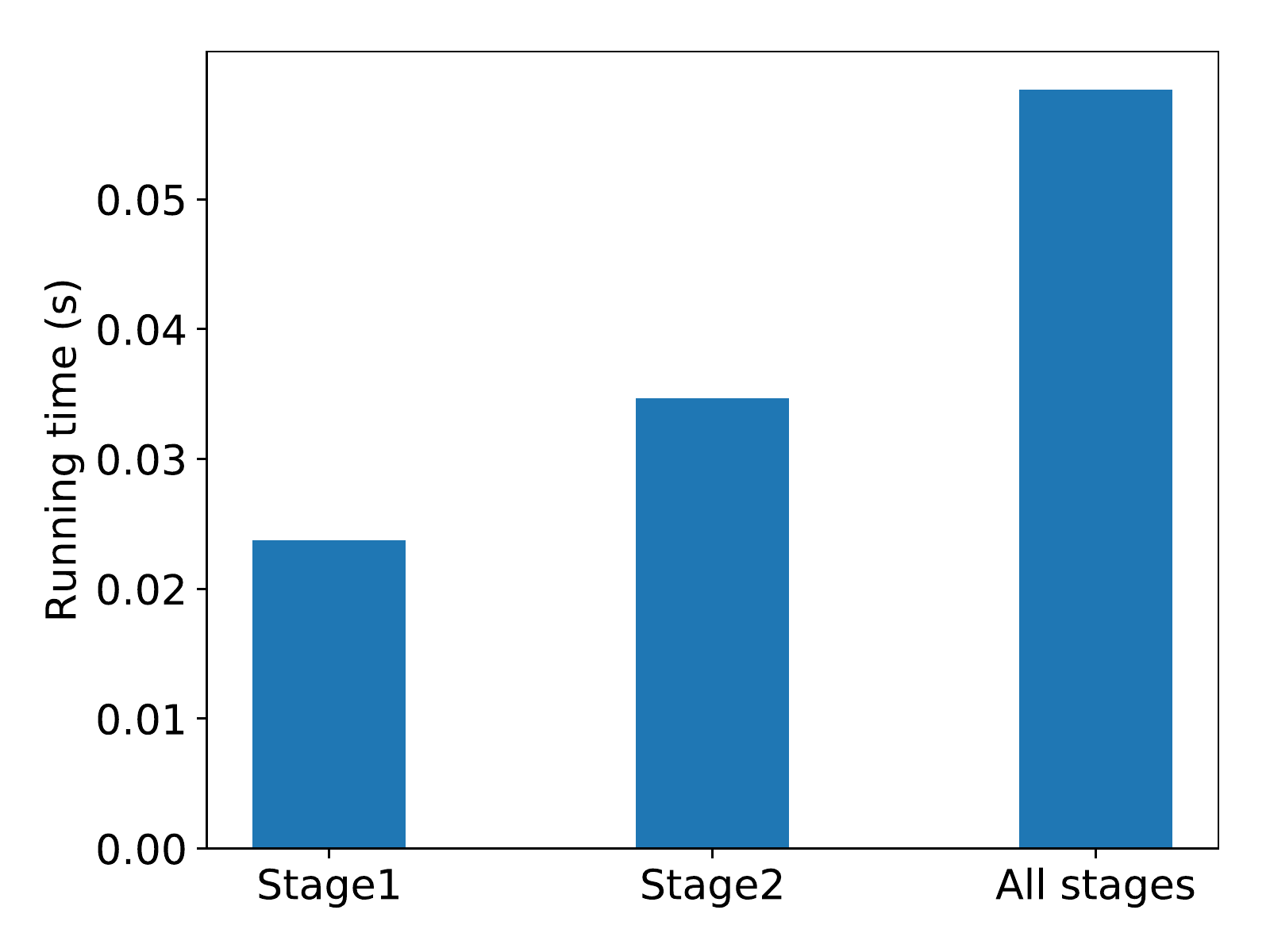}
\caption{}
\label{PPCSS_user_time}
\end{subfigure}\quad 
\begin{subfigure}[b]{0.23\textwidth}
\includegraphics[width=\linewidth]{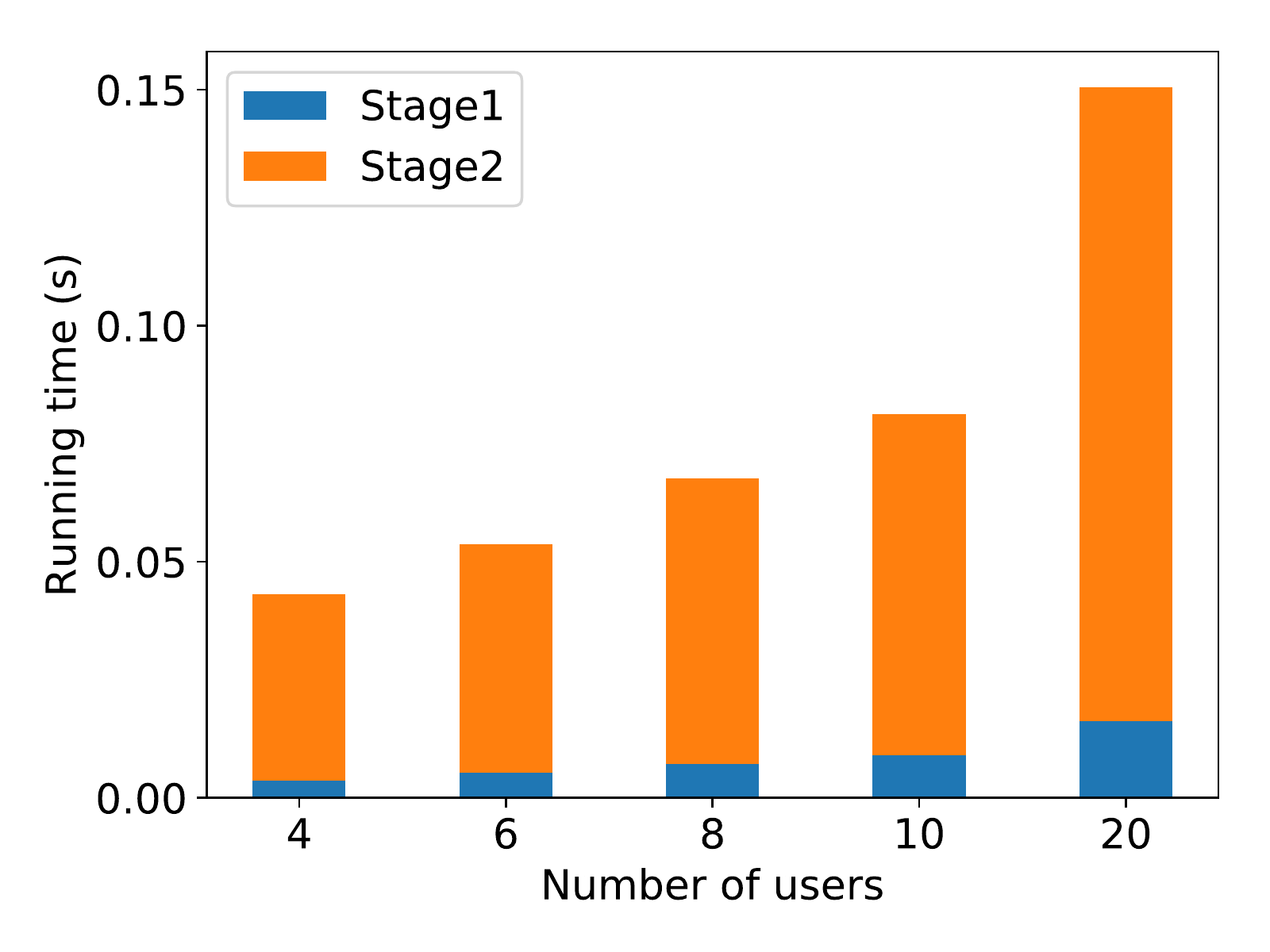}
\caption{}
\label{PPCSS_operator_time}
\end{subfigure}
\caption{PPCSS: (a) Average user running time of different stages. (b) Operator running time vs. number of users.}
\label{fig:PPCSS_time}
\end{figure}

We intentionally ignore the cost of Stage 3 in both PP-UFS and PP-CSS, as the cost-sharing and payment in Stage 3 should be specified by a particular platform for security consideration, such as blockchain.
Therefore, we only focus on the computational cost in the first two stages, which is composed of the cost on each user and the cost on the operator. Here, we set the scaling factor as $100$ to ensure high precision with little effect on the computational time compared with the number of users. Below, we detail the incurred 
computational cost in PP-UFS of Section~\ref{sec:PPUFS}.

On the user side, there are two major processing stages: \textbf{1) Stage 1}: encrypting $b_i^j$ in Stage 1 (1), and decrypting the received ciphertext in Stage 1 (3); \textbf{2) Stage 2}: calculating the ciphertext in Stage 2 (2), and decrypting the received ciphertext in Stage 2 (4). In this experiment, since users can run their stages in parallel, we average the total running time across all users. Fig.~\ref{PPUFS_user_time} illustrates the average running time of each stage, and the total time of two stages. It shows that the total running time is no more than 0.05s, which is well sustainable for the users. All the results in Fig.~\ref{PPUFS_user_time} are average running time across all timeslots derived from 20 users.

On the operator server, there are also two major processing stages: \textbf{1) Stage 1}: multiplying all the received ciphertexts in Stage 1 (2); \textbf{2) Stage 2}: multiplying all the received ciphertexts in Stage 2 (3), and deriving the aggregate schedule of usage $c^j$ in Stage 2 (5). For each timeslot, we evaluate the running time of each stage, and the total running time under different number of users ranging from 4 to 20. 
From Fig.~\ref{PPUFS_operator_time}, we can see that Stage 1 dominates the total running time, because the exponential operation on the ciphertexts in Stage 1 (2) of Section~\ref{sec:PPUFS} costs the majority of running time.

Similarly, Fig.~\ref{PPCSS_user_time} and Fig.~\ref{PPCSS_operator_time} illustrate the incurred computational cost on the user side and on the operator server in PP-CSS of Section~\ref{sec:PPCSS}. Fig.~\ref{PPCSS_user_time} shows that the total average user running time is only 0.057s, which is well within the realm of practicality. Fig.~\ref{PPCSS_operator_time} follows the similar trend as in Fig.~\ref{PPUFS_operator_time}, the only difference is that in PP-CSS, Stage 2 dominates the total running time, because the exponential operation on the ciphertexts in Stage 2 (4) of Section~\ref{sec:PPCSS} occupies the majority of the running time.

However, it should be noted that both the user time given in Fig.~\ref{PPUFS_user_time} and Fig.~\ref{PPCSS_user_time}, and the operator time given in Fig.~\ref{PPUFS_operator_time} and Fig.~\ref{PPCSS_operator_time} are based on a single machine deployed on the user side and the operator server. In practical scenarios, the operator can deploy as many as machines to run in parallel to speed up computation, and the number of machines deployed on the operator should be increased as per the number of users. The users can use more powerful multi-core mobiles to improve efficiency of computation. Therefore, the computational cost is not an obstacle for the implementation of our protocols.

\subsubsection{Communication Cost} 
\

For the complexity of communication, we estimate the communication cost incurred at the first two stages by using the typical 256-bit key size, where a Paillier ciphertext is 1024 bits (128 bytes) and a value in the plaintext message space of Paillier is 1024 bits (128 bytes). Therefore, in PP-UFS as shown in Fig.~\ref{fig:protocol}, the total communication cost of the first two stages between the operator and any user can be estimated as $T_N$ * (1024*4+1024+1024) bits = $T_N$ * 0.75 KB, where $T_N$ is the number of timeslots. Similarly, the first two stages of PP-CFS in Section~\ref{sec:PPCFS} incur the same communication cost as the first two stages of PP-CFS in Section~\ref{sec:PPUFS}. In comparison, for PP-CSS in Section~\ref{sec:PPCSS}, the total communication cost of the first two stages between the operator and any user can be estimated as $T_N$ * (1024*6+1024) bits = $T_N$ * 0.875 KB. Therefore, the communication costs of our protocols are all well within the realm of practicality.

\begin{figure}[!htp]
\centering
  \begin{minipage}[b]{0.23\textwidth}
\includegraphics[width=\linewidth]{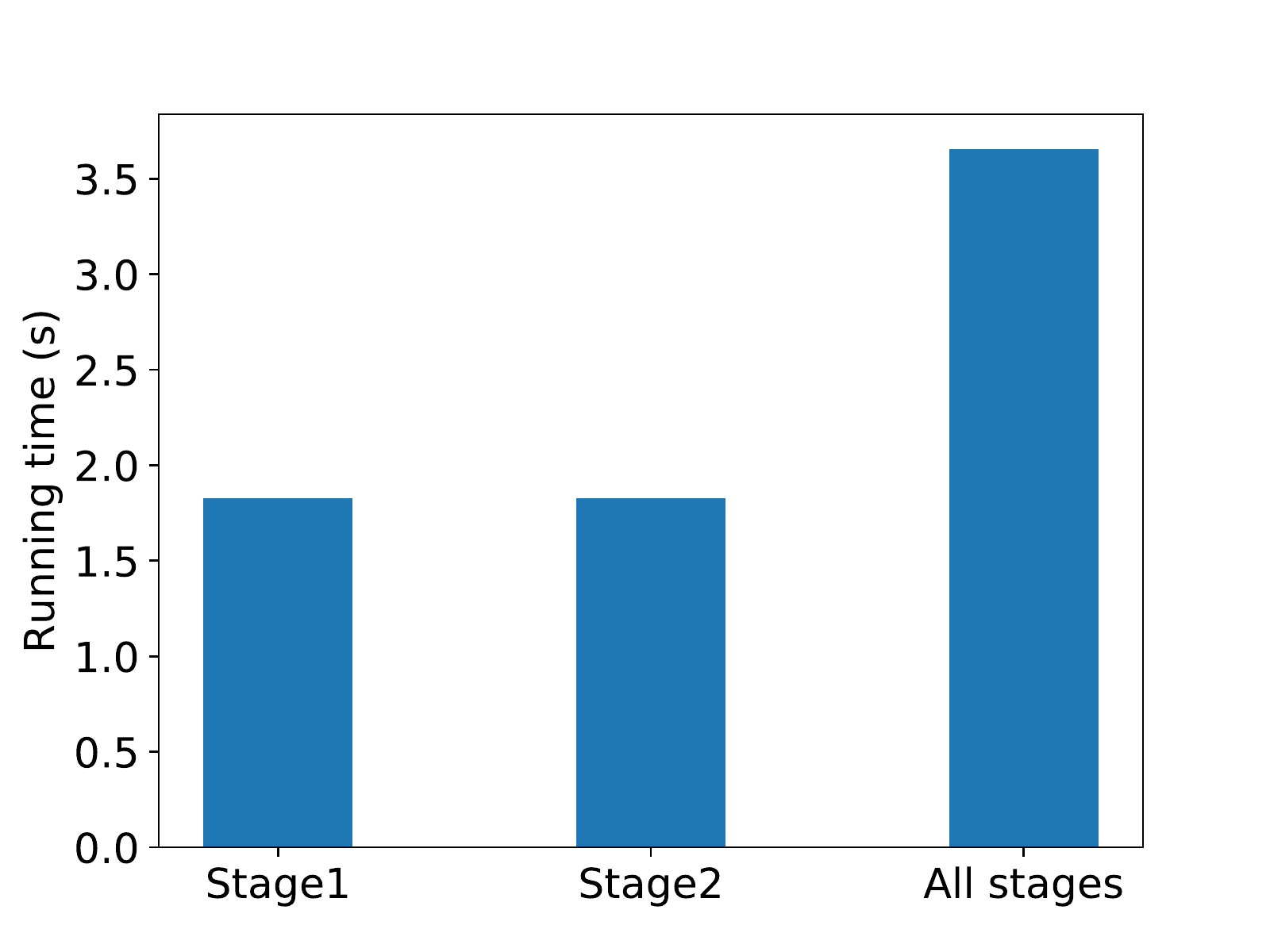}
\caption{Average user running time of PP-UFS using threshold Paillier.}
\label{PPUFS_user_time_threshold}
  \end{minipage}\quad
  \begin{minipage}[b]{0.23\textwidth}
\includegraphics[width=\linewidth]{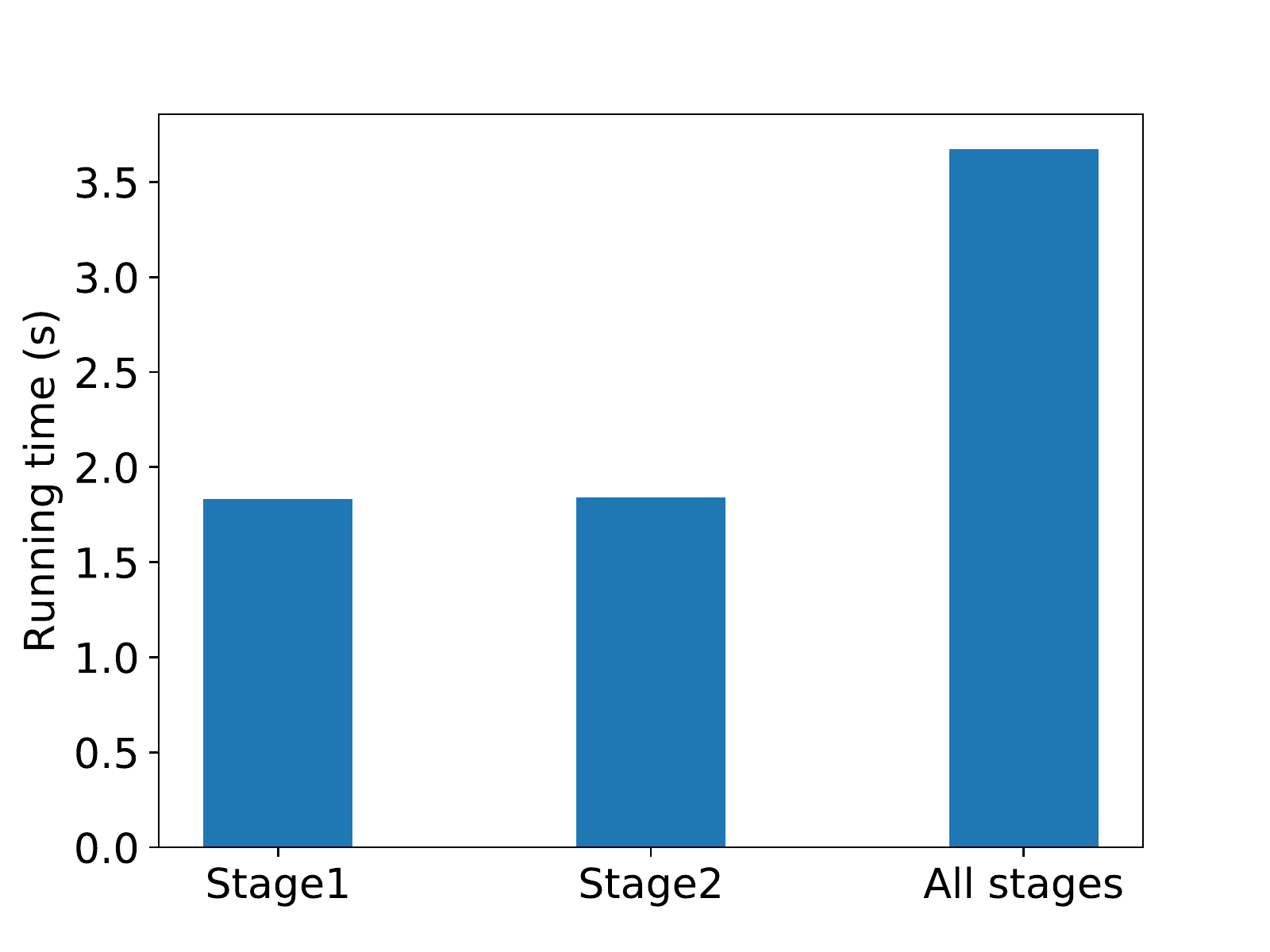}
\caption{Average user running time of PP-CSS using threshold Paillier.}
\label{PPCSS_user_time_threshold}
  \end{minipage}
\end{figure}

\subsubsection{Threshold Paillier Cryptosystems}
\

Fig.~\ref{PPUFS_user_time_threshold} and Fig.~\ref{PPCSS_user_time_threshold} illustrate the incurred computational cost on the user side in PP-UFS and PP-CSS with $(N,N/2)$-threshold Paillier, where $N=20$. In comparison with Fig.~\ref{PPUFS_user_time} and Fig.~\ref{PPCSS_user_time}, it can be observed that threshold Paillier brings additional computational costs but within an acceptable level, while greatly enhancing the security of the system. It should be noted that for $N \leq 3$, $(N,N)$-threshold Paillier should be used instead for security purpose. We remark that the incurred computational costs on the operator server are similar as Fig.~\ref{PPUFS_operator_time} and Fig.~\ref{PPCSS_operator_time}, as the additional computation costs are imposed on the user side rather than on the operator server in threshold Paillier. We also remark that user running time can be significantly reduced through parallel computation with multi-core mobile devices, which we leave as optimization in future work. Note that in threshold Paillier, the communication cost during each decryption is dependent on the number of users $t$ participating in the partial decryption.

\section{Conclusion} \label{sec:concl}

This paper presents the first work of cloud-based privacy-preserving protocols for collaborative consumption. We investigated three real-world problems in collaborative consumption in practical sharing economy applications, including uncapacitated facility sharing, capacitated facility sharing, and communal service sharing. Based on the homomorphic cryptosystems, we propose three novel privacy-preserving protocols to solve these problems. An extensive evaluation study is provided and a proof-of-concept cloud-based system prototype of our protocols is developed. In future, we will extend our cloud-based privacy-preserving protocols to other real-world problems, and devise more robust solutions against malicious users.

\bibliographystyle{IEEEtran}
\bibliography{reference}


\end{sloppypar}
\end{document}